\documentclass[ACS,STIX1COL]{WileyNJD-v2}
\articletype{ORIGINAL ARTICLE}

\usepackage{mathtools}
\mathtoolsset{showonlyrefs=true}
\providecommand{\url}[1]{\texttt{#1}}

\providecommand{\Capitalize}[1]{\uppercase{#1}}
\providecommand{\capitalize}[1]{\expandafter\Capitalize#1}

\providecommand{\bbleds}{eds: }

\providecommand{\bbledn}{ed.}

\providecommand{\bblin}{in}

\providecommand{\bblof}{of}

\providecommand{\bblp}{p}

\providecommand{\bblsecondo}{2nd}

\received{xxx}
\revised{xxx}
\accepted{xxx}

\newcommand{\onlinecite}[1]{\hspace{-1 ex} \nocite{#1}\citenum{#1}} 

\usepackage{bm}

\begin{document}

\title{Time-dependent charged particle stopping in quantum plasmas: testing the G1--G2 scheme for quasi-one-dimensional systems
}

\author[1]{C. Makait}
\author[1]{F. Borges Fajardo}
\author[1]{M. Bonitz}

\authormark{C. Makait \textsc{et al}}

\address[1]{\orgdiv{Institut f\"ur Theoretische Physik und Astrophysik}, \orgname{Christian-Albrechts-Universit\"at zu Kiel}, \orgaddress{\state{Leibnizstra{\ss}e 15, 24098 Kiel}, \country{Germany}}}

\corres{*\email{makait@physik.uni-kiel.de}}

\abstract{Warm dense matter--an exotic, highly compressed state on the boarder between solid and plasma phases is of high current interest, in particular for compact astrophysical objects, high pressure laboratory systems, and inertial confinement fusion. For many applications the interaction of quantum plasmas with energetic particles is crucial. Moreover, often the system is driven far out of equilibrium. In that case, there is high interest in time-dependent simulations to understand the physics, in particular, during thermalization. Recently a novel many-particle technique, the 
G1--G2 scheme was presented [N. Schlünzen et al., Phys. Rev. Lett.  \textbf{124}, 076601 (2020)] which allows for first-principle simulations of the time evolution of interacting quantum systems. Here we apply this scheme to a spatially uniform dense quantum plasma (jellium) and explore its performance. To this end the G1--G2 scheme is transformed into momentum representation, and first results are presented for a quasi-one-dimensional model system.
}


\maketitle

\section{Introduction}\label{sec:1}

Warm dense matter (WDM)--an exotic state on the border of plasma physics and condensed matter physics, e.g.~Refs.~\onlinecite{graziani-book,Fortov2016, moldabekov_pre_18, dornheim_physrep_18}, is currently a very active research field. Among the occurences are the interior of giant planets  [\onlinecite{Militzer_PRE_2021,schlanges-etal.95cpp,bezkrovny_pre_4, vorberger_hydrogen-helium_2007, militzer_massive_2008, redmer_icarus_11,nettelmann_saturn_2013}], brown and white dwarf stars [\onlinecite{saumon_the_role_1992, chabrier_quantum_1993,chabrier_cooling_2000}], and the outer crust of neutron stars [\onlinecite{Haensel,daligault_electronion_2009}]. 
In the laboratory, WDM is being produced via laser or ion beam compression, or with Z-pinches, see Ref.~\onlinecite{falk_2018} for a recent review. 
Aside from dense plasmas, also many condensed matter systems exhibit WDM behaviour -- if they are subject to strong excitation, e.g. by lasers or free electron lasers~[\onlinecite{Ernstorfer1033,PhysRevX.6.021003}].
Among the most important applications is inertial confinement fusion where recently major breakthroughs, including ignition of fusion were reported [\onlinecite{icf-ignition_22}]. Promising fusion relevant results with dense plasmas were also reported in magnetized liner fusion (MagLiF) at Sandia, e.g. Ref.~\onlinecite{maglif_prl_14}.\\

In warm dense matter experiments, collisional heating is an important excitation mechanism.
An example is inverse bremsstrahlung heating in a strong laser field, e.g. [\onlinecite{kremp_99_pre,haberland_01_pre, bonitz_99_cpp}]. Another mechanism is the direct energy transfer from energetic particles to the bulk plasma which is quantified by the stopping power, which has been studied extensively in many fields, including dense plasmas. Direct ion impact has also been proposed as a way to ignite inertial confinement fusion [\onlinecite{roth_prl_01}]. However, reliable theoretical predictions of the energy transfer are still missing.
The reason is that, for warm dense matter, computation of the stopping power faces problems, due to the need to simultaneously take into account electronic quantum effects, moderate to strong Coulomb correlations and finite temperature effects. Quantum effects of electrons are of relevance at low temperature and/or if matter is very highly compressed, such that the temperature is of the order of (or lower than) the Fermi temperature, for a recent overview, see Ref.~[\onlinecite{bonitz_pop_20}]. The most accurate results for thermodynamic properties of warm dense matter, in particular hydrogen, so far, were obtained via first principle computer simulations such as path integral Monte Carlo (PIMC) [\onlinecite{sign_cite,militzer_path_2000,filinov_ppcf_01,filinov_pss_00,pierleoni_cpp19}], however, they are hampered by the fermion sign problem. 
Reliable theoretical data for the electronic component, under thermodynamic equilibrium conditions, have recently become available via a combination of two novel QMC simulations:  configuration PIMC and permutation-blocking PIMC simulations  [\onlinecite{schoof_cpp11,dornheim_physrep_18,filinov_pre15,schoof_prl15,dornheim_njp15}]. However, the stopping power is a dynamical quantity which is not directly accessible to QMC simulations. On the other hand, QMC simulations are able to produce first principle dynamic quantities within linear response theory (LRT), including the dynamic structure factor [\onlinecite{dornheim_prl_18}], the density response,  and dielectric function [\onlinecite{hamann_prb_20}] as well as the plasmon dispersion [\onlinecite{hamann_cpp_20}]. 
One way to apply these results for the stopping power is to make use of QMC data for the local field correction, that were computed in Ref.~[\onlinecite{dornheim_jcp_19-nn}]. This was realized by Moldabekov \textit{et al.} in Ref.~[\onlinecite{moldabekov_pre_20, zhandos_cpp_21}] and yields significant improvements compared to the standard random phase approximation (RPA) results, providing valuable benchmarks for the electronic component of WDM, as long as LRT is applicable. On the other hand, linear response will fail in case of strong driving, where nonlinear effects may become important. The corresponding extensions of QMC simulations beyond LRT were reported in Refs.~[\onlinecite{dornheim_prl_20, dornheim_prr_21, dornheim_cpp22, dornheim_pop_23}].  

On the other hand, LRT for the stopping power will also fail if the system is driven far away from equilibrium, e.g. by fast excitation scenarios. In that case nonequilibrium and non-adiabatic approaches are required. This includes quantum hydrodynamics [\onlinecite{zhandos_pop18, zhandos_cpp17_1d, bonitz_pre_13,zhandos_cpp_19}], Bohmian dynamics [\onlinecite{Lardereaaw1634}] and time-dependent density functional theory (TDDFT)-Ehrenfest simulations [\onlinecite{correa_prl_12,schleife_prb_15,magyar_cpp_16,schluenzen_cpp_18, kononov_nl_21}]. In addition, for dense fully ionized plasmas also quantum kinetic theory simulations of electron relaxation and ion stopping were performed, e.g. Refs. [\onlinecite{gericke_prl11, vorberger_pre_10, kosse-etal.97, graziani_pop_13, grabowski_prl_13}]. Of particular interest in dense plasmas is the role of screening which has been predicted to play a crucial role also for nuclear fusion rates [\onlinecite{ichimaru_rmp_93, langanke_fp_22}]. This requires to solve a quantum kinetic equation with Balescu-Lenard-type collision integrals for dense plasma conditions which has occasionally been attempted, e.g. [\onlinecite{bonitz-etal.96ib2}] and indicated a strong enhancement of the stopping power compared to static screening (Landau equation). However, these results do not yet allow for reliable predictions of the stopping power because they do not simultaneously include strong coupling effects and quantum exchange. Moreover, the Balescu-Lenard equation does not conserve total energy and does not capture the formation of correlations, screening and of plasmons.

In recent years there has been significant progress in the derivation of generalized non-Markovian quantum kinetic equations that overcome these limitations, for an overview, see the text book [\onlinecite{bonitz_qkt}].
Due to the time retardation these equations exhibit an unfavorable cubic (in case of the Keldysh-Kadanoff-Baym equations) or quadratic (in case of the time-diagonal approximation to these equations) scaling with the number of time steps, so numerical solutions pose challenges. Nevertheless, solutions have been reported for dense plasmas and atomic systems [\onlinecite{haberland_01_pre,balzer_pra_10,balzer_pra_10_2}, but only with static screening. Recently, several breakthroughs could be achieved by Schl\"unzen \textit{et al.} [\onlinecite{schluenzen_prl_20,joost_prb_20}]. By eliminating the memory integral the scaling could be reduced to time-linear within the so-called G1--G2 scheme. Moreover, this scaling could be demonstrated for both static and dynamics screening, for the case of correlated electrons in lattice models. Finally, also selfconsistent combination of dynamical screening (polarization of GW diagrams) and strong coupling (ladder diagrams, T-matrix approximation) within the dynamically screened ladder approximation (DSL) was reported for nonequilibrium systems by Joost \textit{et al.} [\onlinecite{joost_prb_22, joost_phd_2022}. Thus, finally a theoretical and computational scheme is available that should allow for predictive nonequilibrium quantum plasma simulations that selfconsistently include quantum, exchange, dynamical screening and strong coupling effects.

In this work, we present the first results of applying the G1--G2 quantum kinetic scheme to dense plasmas. We explore the computational cost of different geometries. As a result of this analysis we conclude that, presently, only quasi-one-dimensional systems can be treated without further simplifications. This is a purely technical limitation which will be overcome with future hardware offering larger computer memory. Therefore, we concentrate on the performance of the G1--G2 scheme for nonadiabatic stopping power simulations in quasi-1D plasmas which may be realized e.g. in quantum wires or in plasmas in a strong magnetic field. For the first implementation of the scheme for dense plasmas we concentrate on the statically screened second Born approximation, deferring dynamical screening (GW) and DSL simulations to future work.\\

The article is organized as follows. In section \ref{s:g1-g2} we present the G1--G2 scheme and apply it to a spatially uniform plasma by introducing the momentum representation. In Sec.~\ref{s:quasi-1d} we introduce the quasi-1D model and compute the Coulomb matrix elements. After this, in Sec.~\ref{s:results} we present our numerical results and present our conclusions and outlook in Sec.~\ref{s:discussion}.

\section{G1--G2 scheme applied to spatially uniform Coulomb systems}\label{s:g1-g2}
\subsection{Hamiltonian and main definitions}
\label{ss:hamiltonian}
Spatially uniform systems are most efficiently described in momentum representation which will be used in the following. We will consider a two-component system consisting of electrons and ions. Let $\hat{a}^{(\dagger)}_{\mathbf{k}\sigma}$ and $\hat{b}^{(\dagger)}_{\mathbf{k}\sigma}$ be the ladder operators of electrons and ions, respectively, known from second quantization theory. Then the Hamiltonian can be written in the form 
\begin{align}
  \hat{H}=\hat{H}_{\text{ee}}+\hat{H}_{\text{ii}}+\hat{H}_{\text{ei}}+\hat{H}_{be}+\hat{H}_{bi}+H_{bb},
\end{align}
($e$: electron, $i$: ion, $b$: background) where $\hat{H}_{ee}$ denotes the pure electron  contribution given by
\begin{align}
    \hat{H}_{\text{ee}}=\sum\limits_{\textbf{k},\sigma}\frac{\hbar^2_{}\textbf{k}^2}{2m_{\text{e}}}
    \hat{a}_{\textbf{k}\sigma}^{\dagger}
    \hat{a}_{\textbf{k}^{}\sigma}+
    \frac{1}{2}\sum\limits_{\textbf{k}\textbf{k}'\textbf{q}\atop \sigma\sigma'}v_{\text{ee}|\textbf{q}|}(t)
    \hat{a}_{\textbf{k}+\textbf{q},\sigma}^{\dagger}
    \hat{a}_{\textbf{k}'-\textbf{q},\sigma'}^{\dagger}
    \hat{a}_{\textbf{k}'\sigma'}^{}
    \hat{a}_{\textbf{k}\sigma}^{}\,,
\label{HEG_hamiltonian2}
\end{align}
and the ionic contribution, $\hat{H}_{ii}$, is written  analogously. The term, $\hat{H}_{ei}$, accounts for the electron-ion interaction and is given by
\begin{equation}
    \hat{H}_{\text{ei}}=
    \sum\limits_{\textbf{k}\textbf{k}'\textbf{q} \atop \sigma\sigma'}v_{\text{ei},|\textbf{q}|}(t)
    \hat{a}_{\textbf{k}^{}+\textbf{q}^{},\sigma}^{\dagger}
    \hat{b}_{\textbf{k}'^{}-\textbf{q}^{},\sigma'}^{\dagger}
    \hat{b}_{\textbf{k}'^{}\sigma'}^{}
    \hat{a}_{\textbf{k}\sigma}^{}
    .\label{hamiltonian_multicomponent_electron_ion}
\end{equation}
The distributions considered in this work are not charge neutral, i.e. we consider systems where one species outnumbers the other. In order to avoid $\mathbf{q}=0$ divergences originating from a net charge of the system, a static background is introduced, which is represented by $\hat{H}_{be}$, $\hat{H}_{bi}$ and $\hat{H}_{bb}$. Their whole effect condenses into three minor but important changes of $\hat{H}_{ee},\hat{H}_{ii},\hat{H}_{ei}:$ the divergent $\mathbf{q}$ term is cancelled out.
The interaction matrix elements, $v_{ee,|\mathbf{q}|}$, $v_{ii,|\mathbf{q}|}$ and $v_{ei,|\mathbf{q}|}$ are the Fourier transforms of the corresponding Coulomb potentials and depend on the dimensionality and geometry of the system. The model geometry used in this article and the resulting expression are given in Sec.~\ref{s:quasi-1d}.
In all simulations charge neutrality will be assumed which amounts to cancellation of the Hartree mean field terms and a cancellation of all contributions with $q=0$.

Equilibrium plasmas can be characterized using a few dimensionless parameters. One-component plasmas are typically characterized by the coupling parameter (Brückner parameter or Wigner-Seitz radius) $r_s$ defined by
\begin{align}
    B(r_sa_B)n=1\,,
\end{align}
with the function $B$ that gives the volume of the ball with radius $r_sa_B$, and the density $n$. In 1D this is given by  $B(r)=2r.$ 
A small $r_s$ value indicates high density and that the system's kinetic energy is larger than its potential energy. The second parameter describing a one-component plasma is 
\begin{align}
    \Theta=\frac{k_BT}{\epsilon_F},
    \nonumber
\end{align}
the quantum degeneracy parameter, where $\epsilon_F=\hbar^2k_F^2/2m$ is the Fermi energy. A large $\Theta$ parameter corresponds to a high temperature and thus little degeneracy.
For a two-component system there exist dimensionless parameters for both components. Here we will concentrate on isothermal  ($T_e=T_i$) systems. In the following, the most important third parameter will be the mass ratio,

\subsection{G1--G2 equations}
Nonequilibrium Green functions (NEGF) open a systematic approach to describe the dynamics of quantum many-body systems in general, e.g. Refs. [\onlinecite{kadanoff-baym,balzer-book}], and quantum plasmas, in particular, e.g. Refs. ~[\onlinecite{green-book,blue-book}]. The central quantity in Green functions theory is the single-particle NEGF, $G^\gtrless_{ij}(t,t')$, which is defined as
\begin{align}
    G^<_{ij}(t,t')=\pm\frac{1}{\mathrm{i}\hbar}\left\langle \hat{a}_{j}^\dagger(t')\,\hat{a}_i(t)\right\rangle,\qquad\qquad G^>_{ij}(t,t')=\frac{1}{\mathrm{i}\hbar}\left\langle \hat{a}_i(t)\,\hat{a}_{j}^\dagger(t')\right\rangle,
    \label{eq:ggtrless-def}
\end{align}
where the expectation value is taken with respect to some initial density matrix $\rho(t_0).$ The equations of motion of $G^\gtrless$ are the two-time Keldysh-Kadanoff-Baym equations (KBE). Here, we will use, instead, only their time-diagonal version as this allows for a dramatic speedup of the simulations. Moreover, 
single-particle observables are straightforwardly computed using the single-particle reduced density matrix $F_{ij}(t)=\pm\mathrm{i}\hbar G^<_{ji}(t,t)$ which involves the time-diagonal NEGF only. Its equation of motion is given by
\begin{align}
    \mathrm{i}\hbar\frac{\text{d}}{\text{d}t}G^\gtrless_{ij}(t,t)-\sum\limits_{k}\left[h^\text{HF}_{ik}(t)\,G^\gtrless_{kj}(t,t)-G^\gtrless_{ik}(t,t)\,h^\text{HF}_{kj}(t)\right]=I_{ij}(t)+\left[I_{ji}(t)\right]^*,
\end{align}
with the Hartree--Fock Hamiltonian and the (anti-)symmetrized pair potential, 
\begin{align}
    h^\text{HF}_{ij}(t)=h^{(0)}_{ij}\pm\mathrm{i}\hbar\sum\limits_{kl}w^{\pm}_{ikjl}(t)\,G^<_{lk}(t,t),\qquad \qquad w^\pm_{ijkl}=w_{ijkl}\pm w_{jikl}.
\end{align}
The collision integral $I_{ij}(t)$ is given by
\begin{align}
    I_{ij}(t)=\sum\limits_k\int_{t_0}^t\text{d}\bar{t}\,\left[\Sigma^>_{ik}(t,\bar{t})\,G^<_{kj}(\bar{t},t)-\Sigma^<_{ik}(t,\bar{t})\,G^>_{kj}(\bar{t},t)\right]\,,\label{eq:CollIntMemory}
\end{align}
where $\Sigma$ is the correlation part of the many-body selfenergy, a functional of $G$ that includes all many-body effects beyond Hartree-Fock. Note that the collision integral involves also time-off-diagonal components of the Green functions and selfenergies. They are approximated in well-known manner using the Generalized Kadanoff--Baym Ansatz (GKBA) [\onlinecite{lipavski_prb_86}] with Hartree--Fock propagators (HF-GKBA), for details see Refs.~
[\onlinecite{balzer-book,hermanns_prb14}]. 

Recently, the HF-GKBA was reformulated in a time-local way leading to coupled equations for the single-particle Green function on the time diagonal, $G^\gtrless_{ij}(t)\vcentcolon=G^\gtrless_{ij}(t,t)$,  and the correlation part of the two-particle Green function, $\mathcal{G}$, [\onlinecite{schluenzen_prl_20,joost_prb_20}] which was named ``G1--G2 scheme''. In that scheme, the time-nonlocal collision integral is replaced by the local expression
\begin{align}
    I_{ij}(t)=\pm\mathrm{i}\hbar\sum\limits_{klm}v_{imkl}(t)\,\mathcal{G}_{kljm}(t)\,,
\end{align}
where $\mathcal{G}$ depends on the chosen selfenergy approximation.
The equation of motion of $\mathcal{G}$, within the  dynamically screened ladder approximation (DSL, the only approximation is the neglect of three-particle correlations), is given by [\onlinecite{joost_prb_22}] 
\begin{align}
    \mathrm{i}\hbar\frac{\text{d}}{\text{d}t}\mathcal{G}_{ijkl}(t)-\left[h^{\text{HF},(2)}(t),\mathcal{G}(t)\right]_{ijkl}=\Psi^{\pm}_{ijkl}(t)+\Pi_{ijkl}(t) + \Lambda^\text{ph}_{jikl}(t)+\Lambda^\text{pp}_{ijkl}(t)\,,
\label{eq:g2-equation}
\end{align}
with the two-particle Hartree-Fock term
\begin{align}
    \left[h^{\text{HF},(2)}(t),\mathcal{G}(t)\right]_{ijkl}=\sum\limits_{m}\left[h^\text{HF}_{im}(t)\,\mathcal{G}_{mjkl}(t)+h^\text{HF}_{jm}(t)\,\mathcal{G}_{imkl}(t)-\mathcal{G}_{ijml}(t)\,h^\text{HF}_{mk}(t)-\mathcal{G}_{ijkm}(t)\,h^\text{HF}_{ml}(t)\right],
\end{align}
and the (anti-)symmetrized source term
\begin{align}
    \Psi^\pm_{ijkl}(t)=(\mathrm{i}\hbar)^2\sum\limits_{pqrs}\left[G^>_{ip}(t)\,G^>_{jq}(t)\,w^\pm_{pqrs}(t)\,G^<_{rk}(t)\,G^<_{sl}(t)-G^<_{ip}(t)\,G^<_{jq}(t)\,w^\pm_{pqrs}(t)\,G^>_{rk}(t)\,G^>_{sl}(t)\right]\,.
\end{align}
If, on the r.h.s. of Eq.~\eqref{eq:g2-equation}, only $\Psi^\pm$ is retained  ($\Lambda=\Pi=0$), this corresponds to the static  second order Born approximation (SOA) which leads to a non-Markovian generalization of the Landau equation of plasma physics [\onlinecite{bonitz_qkt}] (which in the present reformulation is time-local).
 On the other hand, taking additionally the term $\Pi_{ijkl}(t)$ into account, leads to the nonequilibrium $GW$ approximation [\onlinecite{joost_prb_22}]. Here, the polarization term is given by
\begin{align}
    \Pi_{ijkl}(t)=\pi_{ijkl}(t)-\left[\pi_{lkji}(t)\right]^*,\qquad\text{where}\qquad \pi_{ijkl}(t)=(\mathrm{i}\hbar)^2\sum\limits_{pqrs}(\pm)_j\mathcal{G}_{ipkq}(t)\,w_{sqrp}(t)\left[G^>_{js}(t)\,G^<_{rl}(t)-G^<_{js}(t)\,G^>_{rl}(t)\right],
\end{align}
and $(\pm)_j$ is the sign of the particle species occupying state $j$. This approximation is the non-Markovian generalization of of the Balescu-Lenard kinetic equation. 
For completeness, we also give the T-matrix contributions in the particle-particle (pp) and particle-hole (ph) channels which are associated with the $\Lambda$ terms:
\begin{align}
    \Lambda^\text{ph}_{ijkl}(t)=\lambda^\text{ph}_{ijkl}(t)-\left[\lambda^\text{ph}_{klij}(t)\right]^*,\qquad\text{where}\qquad \lambda^\text{ph}_{ijkl}(t)=(\mathrm{i}\hbar)^2\sum\limits_{pqrs}\mathcal{G}_{qjkp}(t)\,w_{rpqs}(t)\left[G^>_{ir}(t)\,G^<_{sl}(t)-G^<_{ir}(t)\,G^>_{sl}(t)\right],\\
\nonumber
    \Lambda^\text{pp}_{ijkl}(t)=\lambda^\text{pp}_{ijkl}(t)-\left[\lambda^\text{pp}_{klij}\right]^*,\qquad\text{where}\qquad \lambda^\text{pp}_{ijkl}(t)=(\mathrm{i}\hbar)^2\sum\limits_{pqrs}\left[G^>_{ir}(t)\,G^>_{js}(t)-G^<_{ir}(t)\,G^<_{js}(t)\right]w_{rspq}(t)\,\mathcal{G}_{pqkl}(t)\,.
    \nonumber
\end{align}
The advantage of the G1--G2 scheme is that all these approximations can be treated with comparable effort. This offers the opportunity to selfconsistently treat dynamical screening, strong Coulomb correlations, bound states such as atoms or excitons, as well as the buildup of correlations and screening, as was shown for Hubbard-type lattice models, cf. Ref.~[\onlinecite{joost_prb_22}]

In this article we present the first application of the G1--G2 scheme to dense plasmas. To explore the specifics of this problem, we will concentrate on the Second Born approximation deferring improved approximations to future work.

\subsection{Momentum representation of the G1--G2 equations}\label{ss:g1g2-momentum}
The application of the G1--G2 scheme to uniform systems is suitably done in momentum representation. In our case, where different particle species are included, we define a basis whose states are defined by three quantum numbers: the momentum vector $\mathbf{p}$, the spin projection $\sigma$ and the particle species index $\alpha$. In this basis, the single-particle NEGF $G$ is diagonal in the spin and the particle species index, and the interaction potential (screened Coulomb) does neither change the spin nor the species of the interacting particles. Thus in the following, since they evoke the same structures in the equations, greek indices represent both, the spin and the species of the particle, and sums are interpreted as sums over all spin states and particle species.

Then, the single-particle Green function is of the form 
\begin{align}
 G^\gtrless_{\mathbf{p}\alpha,\mathbf{p}'\alpha'}(t)=\vcentcolon G^\gtrless_{\mathbf{p}\alpha}(t)\delta_{\mathbf{p},\mathbf{p}'}\delta_{\alpha,\alpha'}\,,
\end{align}
and the pair interaction takes the form 
\begin{align}
w_{\mathbf{p}_1\alpha_1,\mathbf{p}_2\alpha_2,\mathbf{p}_3\alpha_3,\mathbf{p}_4\alpha_4}=\vcentcolon w_{|\mathbf{p}_1-\mathbf{p}_3|}^{\alpha_1,\alpha_2}\delta_{\alpha_1,\alpha_3}\delta_{\alpha_2,\alpha_4}\delta_{\mathbf{p}_1+\mathbf{p}_2,\mathbf{p}_3+\mathbf{p}_4}. \label{eq:interactForm}   
\end{align}
The species dependence of $w^{\alpha_1,\alpha_2}_{|\mathbf{q}|}$ for Coulomb-like interactions factorizes into $w^{\alpha_1,\alpha_2}_{|\mathbf{q}|}=w_{|\mathbf{q}|}Z_{\alpha_1}Z_{\alpha_2},$ where $Z_{\alpha_{1}},Z_{\alpha_{2}}$ are the charge numbers of the particles. Spatial homogeneity and the momentum-conserving structure of $w$ induce the following structure for $\mathcal{G}:$
\begin{align}
    \mathcal{G}_{\mathbf{k}\alpha,\mathbf{p}\beta,\mathbf{k}'\alpha',\mathbf{p}'\beta'}(t)=\vcentcolon\mathcal{G}^{\alpha\beta}_{\mathbf{k}',\mathbf{p}',\mathbf{k}'-\mathbf{k}}(t)\delta_{\alpha\alpha'}\delta_{\beta\beta'}\delta_{\mathbf{k}+\mathbf{p},\mathbf{k}'+\mathbf{p}'},\quad\text{where we denoted}\quad \mathcal{G}^{\alpha\beta}_{\mathbf{kpq}}(t)=\mathcal{G}_{\mathbf{k}-\mathbf{q},\alpha;\mathbf{p}+\mathbf{q},\beta;\mathbf{k},\alpha;\mathbf{p},\beta}(t)\label{eq:G2StructureMomRep}\,.
\end{align}
Due to these many Kronecker deltas the G1--G2 equations become very compact. The single-particle equation is given by
\begin{align}
    \mathrm{i}\hbar\frac{\text{d}}{\mathrm{d}t}G^\gtrless_{\mathbf{p}\alpha}(t)=I_{\mathbf{p}\alpha}(t)+I_{\mathbf{p}\alpha}^\dagger(t),\qquad I_{\mathbf{p}\alpha}(t)=\pm\mathrm{i}\hbar  Z_\alpha\sum\limits_{\mathbf{kq},\beta}Z_\beta w_{\mathbf{q}}(t)\,\mathcal{G}^{\beta\alpha}_{\mathbf{kpq}}(t),
\end{align}
where the commutator involving the single-particle hamiltonian vanishes, due to diagonality of the operands. The various terms in the $\mathcal{G}$ equation within the GW approximation are\footnote{The usual GW approximation uses the non-antisymmetrized source term. We give here the antisymmetrized variant since it appears in the SOA and TMA},
\begin{align}
    \mathrm{i}\hbar \frac{\text{d}}{\text{d}t}\mathcal{G}^{\alpha\beta}_{\mathbf{kpq}}(t)-\left[h^{\text{HF},(2)}(t),\mathcal{G}(t)\right]_{\mathbf{kpq}}^{\alpha\beta}&=\Psi^{\pm,\alpha\beta}_{\mathbf{kpq}}(t)+\Pi^{\alpha\beta}_{\mathbf{kpq}}(t)\,,
\end{align}
with the definitions
\begin{align}
    \left[h^{\text{HF},(2)}(t),\mathcal{G}(t)\right]_\mathbf{kpq}^{\alpha\beta}&=\mathcal{G}^{\alpha\beta}_{\mathbf{kpq}}(t)\left(h^\text{HF}_{\mathbf{k}-\mathbf{q},\alpha}(t)+h^\text{HF}_{\mathbf{p}+\mathbf{q},\beta}(t)-h^\text{HF}_{\mathbf{k},\alpha}(t)-h^\text{HF}_{\mathbf{p},\beta}(t)\right)\,,
    \label{eq:2pCommutatorMomRep}
\end{align}
and
\begin{align}\nonumber
    \Psi^{\pm,\alpha\beta}_{\mathbf{kpq}}(t)&=(\mathrm{i}\hbar)^2\left[v_{|\mathbf{q}|}^{\alpha\beta}\pm \delta_{\alpha\beta}v^{\alpha\alpha}_{|\mathbf{k}-\mathbf{p}-\mathbf{q}|}(t)\right]\left(G^>_{\mathbf{k}-\mathbf{q},\alpha}(t)\,G^>_{\mathbf{p}+\mathbf{q},\beta}(t)\,G^<_{\mathbf{k},\alpha}(t)\,G^<_{\mathbf{p},\beta}(t)-G^<_{\mathbf{k}-\mathbf{q},\alpha}(t)\,G^<_{\mathbf{p}+\mathbf{q},\beta}(t)\,G^>_{\mathbf{k},\alpha}(t)\,G^>_{\mathbf{p},\beta}(t)\right)\,,\\
    \Pi^{\alpha\beta}_{\mathbf{kpq}}(t)&=\pi_{\mathbf{kpq}}^{\alpha\beta}(t)-\left[\pi_{\mathbf{p}+\mathbf{q},\mathbf{k}-\mathbf{q},\mathbf{q}}^{\beta\alpha}(t)\right]^*,\quad\text{where}\quad \pi_{\mathbf{kpq}}^{\alpha\beta}=(\pm)_\beta(\mathrm{i}\hbar)^2\left[G^>_{\mathbf{p}+\mathbf{q},\beta}(t)\,G^<_{\mathbf{p},\beta}(t)-G^<_{\mathbf{p}+\mathbf{q},\beta}(t)\,G^>_{\mathbf{p},\beta}(t)\right]\sum\limits_{\mathbf{p}'\gamma}v_{|\mathbf{q}|}^{\alpha\gamma}(t)\,\mathcal{G}^{\alpha\gamma}_{\mathbf{k}\mathbf{p}'\mathbf{q}}(t)\,.
    \nonumber
\end{align}

Due to spatial homogeneity, the Hartree--Fock Hamiltonian in Eq. \eqref{eq:2pCommutatorMomRep} contains only the Fock term (the Hartree-term vanishes),
\begin{align}
    h^\text{HF}_{\mathbf{p}\sigma}(t)=\frac{\mathbf{p}^2}{2m_\sigma}+\mathrm{i}\hbar \sum\limits_{\mathbf{q}}v^{\sigma\sigma}_{|\mathbf{p}-\mathbf{q}|}(t)\,G^<_{\mathbf{q},\sigma}(t).\label{eq:HFHamiltonianMom}
\end{align}
Note that this Hamiltonian is purely real. It is derived from the two-time KBE by applying the Hartree-Fock GKBA (HF-GKBA) which neglects (in the propagators) correlation and finite quasiparticle life time  effects. Correspondingly, the single-particle spectral function is a delta function, $a^{\rm HF}(\textbf{p},\sigma;\omega,t)=2\pi\delta\left[\hbar\omega - h^\text{HF}_{\mathbf{p}\sigma}(t)\right] $. Alternatively, to restore quasiparticle damping effects approximately, we may add a small time- and momentum-independent damping to the single-particle energy,
\begin{align}
    h^\text{LHF}_{\mathbf{p}\sigma}(t)=h^\text{HF}_{\mathbf{p}\sigma}(t) + \mathrm{i}\hbar\gamma\,,
    \label{eq:hhf-damped}
\end{align}
which assures a finite quasi-particle life time of the order of $1/\gamma$. We will call this approximation Lorentzian HF-GKBA (LHF-GKBA) because it gives rise to a Lorentzian spectral function,
\begin{align}
    a^{\rm LHF}(\textbf{p},\sigma;\omega,t)=\frac{2\hbar\gamma}{\left[\hbar\omega - h^\text{HF}_{\mathbf{p}\sigma}(t)\right]^2 + (\hbar\gamma)^2}\,,
\end{align}
which approaches the quasiparticle spectral function $a^{\rm HF}$ when $\gamma\to 0$. While the LHF-GKBA violates total energy conservation and breaks time reversibility, see Refs.~ \onlinecite{scharnke_jmp17,bonitz_cpp18}, this effect is small, as long as $\gamma$ is small, for details, see Ref.~\onlinecite{bonitz_qkt}. 
The behavior of the Lorentzian HF-GKBA has been tested in detail against two-time KBE simulations by Bonitz et al. in Ref.~\onlinecite{bonitz-etal.99epjb} where also estimates for $\gamma$ for a uniform electron gas are  provided. 

Below, in Sec.~\ref{s:results}, we will report G1--G2 results that use, both, the HF-GKBA and the LHF-GKBA. There we will observe that HF-GKBA simulations may become unreliable for long simulation times and that this effect can be cured by resorting to the LHF-GKBA instead.
Finally, in the thermodynamic limit, momentum summations are replaced by an integral,  $\sum\limits_{\mathbf{p}}\longrightarrow \int\frac{d\mathbf{p}}{(2\pi\hbar)^d}$, where $d$ is the dimension of the system.

\subsection{Observables}
Expectation values of $s$-particle observables $\hat{A}^s$ can be computed from the $s$-particle reduced density operator $\hat{F}^s$ by [\onlinecite{bonitz_qkt}]
\begin{align}
    \left\langle \hat{A}^{(s)}\right\rangle=\frac{1}{s!}\text{Tr}_{1\dots s}\,\hat{F}^{(s)}\hat{A}^{(s)}\,.
\end{align}
The G1--G2 scheme gives direct access to the one- and two-particle density matrices (the matrix representations of the density operators),
\begin{align}
    {F}^{(1)}_{ij}(t)&=\pm\mathrm{i}\hbar G^<_{ij}(t)\,,\\
    {F}^{(2)}_{ijkl}(t)&=(\mathrm{i}\hbar)^2\left({G}^H_{ijkl}(t)\pm{G}^F_{ijkl}(t)+\mathcal{G}_{ijkl}(t)\right)=\left(\mathrm{i}\hbar\right)^2\left(G^<_{ik}(t)\,G^<_{jl}(t)\pm G^<_{il}(t)\,G^<_{jk}(t)+\mathcal{G}_{ijkl}(t)\right),\label{eq:2pRDM}
\end{align}
($H$: Hartree, $F$: Fock) and thus to one- and two-particle observables. Because of the diagonality of $G$ in momentum representation, expectation values of single-particle observables can be computed from
\begin{align}
    \left\langle \hat{A}^{(1)}\right\rangle(t)= \mathrm{i}\hbar \sum\limits_{\alpha}(\pm)_\alpha\int\frac{\mathrm{d}\mathbf{p}}{(2\pi\hbar)^d} A^{(1)}_{\mathbf{p}\alpha}G^<_{\mathbf{p}\alpha}(t).
\end{align}
In addition, we have access to the observable of each species. For example, the particle number density, momentum density, and kinetic energy density for spin/species component ``a'' are given by
\begin{align}
    \left\langle \hat{n}_\alpha\right\rangle(t)&=\mathrm{i}\hbar(\pm)_\alpha \int\frac{\mathrm{d}\mathbf{p}}{(2\pi\hbar)^d}G^<_{\mathbf{p}\tilde{\alpha}}(t)\,,\\
    \left\langle \hat{\mathbf{p}}_\alpha\right\rangle(t)&=\mathrm{i}\hbar(\pm)_\alpha\int\frac{\mathrm{d}\mathbf{p}}{(2\pi\hbar)^d}\mathbf{p}\,G^<_{\mathbf{p}\alpha}(t)\,,\\
    \left\langle \hat{T}_\alpha\right\rangle(t)&=\mathrm{i}\hbar(\pm)_\alpha\int\frac{\mathrm{d}\mathbf{p}}{(2\pi\hbar)^d}\frac{\mathbf{p}^2}{2m_\alpha}\,G^<_{\mathbf{p}\alpha}(t)\,.
\end{align}
The interaction energy, as a two-particle observable, is computed from the two-particle Green function, cf. Eq. \eqref{eq:2pRDM}. The Hartree contribution vanishes due to charge neutrality, as noted above. The Fock-exchange energy is computed from $G^<$ by 
\begin{align}
    \left\langle \hat{w}\right\rangle^F(t)&=\frac{1}{2}(\mathrm{i}\hbar)^2 \sum\limits_\alpha\int\frac{\mathrm{d}\mathbf{k}}{(2\pi\hbar)^d}\frac{\mathrm{d}\mathbf{p}}{(2\pi\hbar)^d}Z_\alpha^2\,G^<_{\mathbf{k}\alpha}(t)\,G^<_{\mathbf{p}\alpha}(t)\,w_{\mathbf{k}-\mathbf{
    p}}=\frac{1}{2} \mathrm{i}\hbar\sum\limits_\alpha (\pm)_\alpha \int \frac{\mathrm{d}\mathbf{k}}{(2\pi\hbar)^d} h^\text{HF}_{\mathbf{k}\alpha}(t)\,G^<_{\mathbf{k}\alpha}(t)-\frac{1}{2}\left\langle\hat{T}\right\rangle(t)\,,
\end{align}
and the final interaction contribution, the correlation part ($c:$ correlation), is computed from $\mathcal{G},$
\begin{align}
    \left\langle \hat{w}\right\rangle^c(t)=\frac{1}{2}(\mathrm{i}\hbar)^2\sum\limits_{\alpha\beta}Z_\alpha Z_\beta\int\frac{\mathrm{d}\mathbf{k}}{(2\pi\hbar)^d}\frac{\mathrm{d}\mathbf{p}}{(2\pi\hbar)^d}\frac{\mathrm{d}\mathbf{q}}{(2\pi\hbar)^d}\mathcal{G}^{\alpha\beta}_{\mathbf{kpq}}(t)\,w_\mathbf{q}=\frac{1}{2}\mathrm{i}\hbar\sum\limits_{\alpha}(\pm)_\alpha \int\frac{\mathrm{d}\mathbf{p}}{(2\pi\hbar)^d}I_{\mathbf{p}\alpha}(t)\,.
\end{align}

\subsection{Scaling of the numerical effort of the G1--G2 scheme for jellium}\label{ss:scaling}
Here we analyze the computational effort required to solve the quantum kinetic equations of motion for jellium, extending the estimates that were presented in Ref.~[\onlinecite{schluenzen_prl_20}]. We present the CPU and RAM scalings for the second order approximation (SOA),  comparison to the standard HF-GKBA with the memory integral formulation. There, the direct (d) term of the two SOA contributions can be computed efficiently using Fourier transforms, whereas the exchange (x) term is a lot more costly since it is not of convolution structure. Hence we distinguish between the direct only and the full second order approximation in Table  \ref{tab:SOAScalings}. Note that, in contrast, for the G1--G2 scheme both terms require the same effort. 
\begin{table}[hbt]
    \centering
\begin{tabular}{|cc|ccc|cc|}
\toprule
    \multicolumn{2}{|c|}{\multirow{2}{*}{SOA}} &\multicolumn{3}{c|}{CPU time} & \multicolumn{2}{c|}{RAM} \\
    & &  GKBA d. & GKBA d. + x. & G1--G2 & GKBA & G1--G2\\
\hline
    1D & & $\mathcal{O}(N_t^2N_x\ln{N_x})$ & $\mathcal{O}(N_t^2 N_x^3)$ & $\mathcal{O}(N_tN_x^3)$ & $\mathcal{O}(N_tN_x)$ & $\mathcal{O}(N_x^3)$\\
\hline
    \multirow{2}{*}{2D} & isotropic   & $\mathcal{O}(N_t^2N_x\ln N_x)$ & $\mathcal{O}(N_t^2 N_x^5)$ & $\mathcal{O}(N_tN_x^5)$ & $\mathcal{O}(N_tN_x)$ & $\mathcal{O}(N_x^5)$ \\
                        & anisotropic & $\mathcal{O}(N_t^2N_x^2\ln N_x)$ & $\mathcal{O}(N_t^2N_x^6)$ & $\mathcal{O}(N_tN_x^6)$ & $\mathcal{O}(N_tN_x^2)$ & $\mathcal{O}(N_x^6)$\\
\hline
    \multirow{3}{*}{3D} & isotropic   & $\mathcal{O}(N_t^2N_x\ln{N_x})$ & $\mathcal{O}(N_t^2N_x^6)$ & $\mathcal{O}(N_tN_x^6)$ & $\mathcal{O}(N_tN_x)$ & $\mathcal{O}(N_x^6)$ \\
                        & cylindric   & $\mathcal{O}(N_t^2N_x^2\ln N_x)$ & $\mathcal{O}(N_t^2N_x^8)$ & $\mathcal{O}(N_tN_x^8)$ & $\mathcal{O}(N_tN_x^2)$ & $\mathcal{O}(N_x^8)$ \\
                        & anisotropic & $\mathcal{O}(N_t^2N_x^3\ln{N_x})$ & $\mathcal{O}(N_t^2N_x^9)$ & $\mathcal{O}(N_tN_x^9)$ & $\mathcal{O}(N_tN_x^3)$ & $\mathcal{O}(N_x^9)$ \\
\bottomrule
\end{tabular}
\caption{Numerical scalings of the CPU time and RAM consumption, for SOA jellium simulations in different dimensions and symmetries. `GKBA' denotes the HF-GKBA within the standard non-Markovian formalism, and `d.' (`x') denotes the direct (exchange) term. G1--G2 scalings for both types are identical. 'GKBA d.' CPU scalings are based on Fast Fourier Transform techniques used for the efficient computation of convolutions. Scalings are based on a Cartesian grid with $N_x$ grid points per axis. The number of time steps is denoted $N_t$. For a typical estimate of RAM and CPU time we refer to Fig.~\ref{fig:ConvergenceTest}.}
    \label{tab:SOAScalings}
\end{table}
In the table the number of time steps and grid points (cartesian, per axis) are denoted by $N_t$ and $N_x$, respectively. While it confirms the known advantage of the G1--G2 scheme in its linear scaling with $N_t$, it is immediately clear, that the scaling with $N_x$ is very unfavorable. The reason is that, in contrast to the standard GKBA, this scheme has to store the current expression of the two-particle Green function $\mathcal{G}$ which has three vector indices (the fourth is eliminated due to spatial homogeneity). Thus, in an anisotropic $d$-dimensional system the matrix $\mathcal{G}$ has $N^{3d}_x$ discrete elements. To resolve the nonequilibrium momentum distribution, $N_x$ has to be of the order of $100$, cf. Sec.~\ref{ss:tests}, thus $\mathcal{G}$ has on the order of $100^{3d}$ complex elements. It is clear that, on current hardware, already an anisotropic $d=2$ situation is practically not feasible.

\section{Quasi-One-Dimensional Model plasma}\label{s:quasi-1d}

\subsection{Pair potential}
\begin{figure}[t] 
\centering
\includegraphics[width=0.5\textwidth]{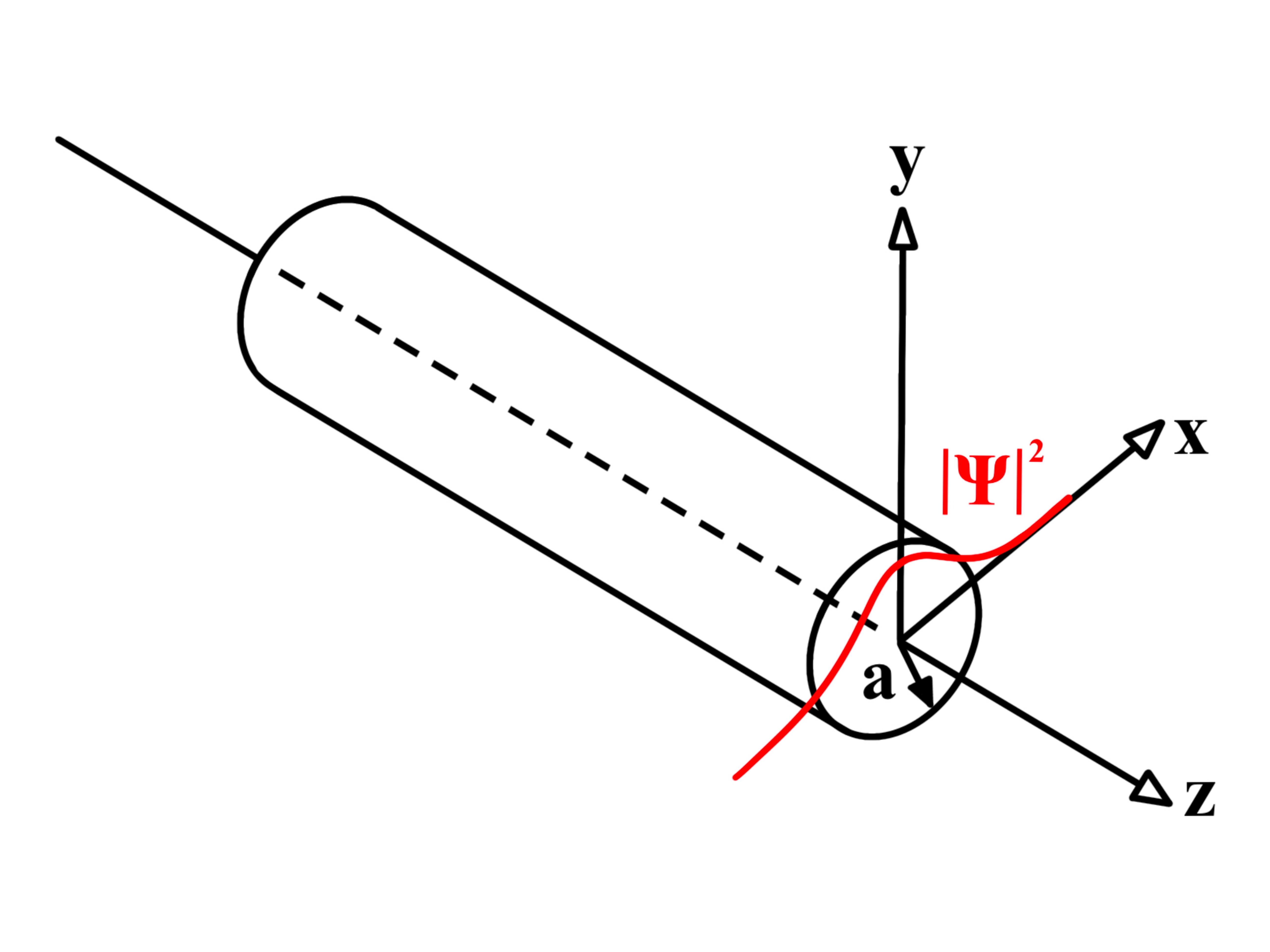}
\caption{Geometry of the quasi-1D plasma. A harmonic confinement is applied, e.g. due to a strong magnetic field, giving rise to a finite transverse extension $a$ of the states.}
\label{Fig:CylinderModel}
\end{figure}
Following the derivations of Ref.~[\onlinecite{giuliani_quantum_2005}], we consider a cylinder-symmetric system (axes $r_\parallel,r_\perp$)  that is radially confined by a harmonic potential
\begin{align}
   V(r_\perp)=\frac{1}{2}m\omega^2r_\perp^2.
   \nonumber
\end{align}
The eigenstates of the corresponding Hamiltonian are products of plane waves in $r_\parallel$-direction, with 2D harmonic oscillator states in the perpendicular direction. If the thermal energy is small compared to the energy spacing of the oscillator states, i.e. $\hbar\omega \gg k_B T$, the occupation of excited oscillator states is negligible, and the wavefunctions have the form
\begin{align}
   \langle\mathbf{r}\alpha'|\mathbf{k}\alpha\rangle = \frac{1}{\sqrt{L}}\left(\frac{2}{\pi a^2}\right)^\frac{1}{2}\exp{\left(-\frac{r_\perp^2}{a^2}\right)}\exp{\left(\mathrm{i}r_\parallel\cdot k\right)}\delta_{\alpha\alpha'},
   \label{eq:1p-states}
\end{align}
cf. Fig.~\ref{Fig:CylinderModel}. Here, we have introduced the width $a$ of the Gaussian groundstate, given by $a^2=2\hbar/m\omega.$

\begin{figure}[t] 
\centering
\includegraphics[width=0.9\linewidth]{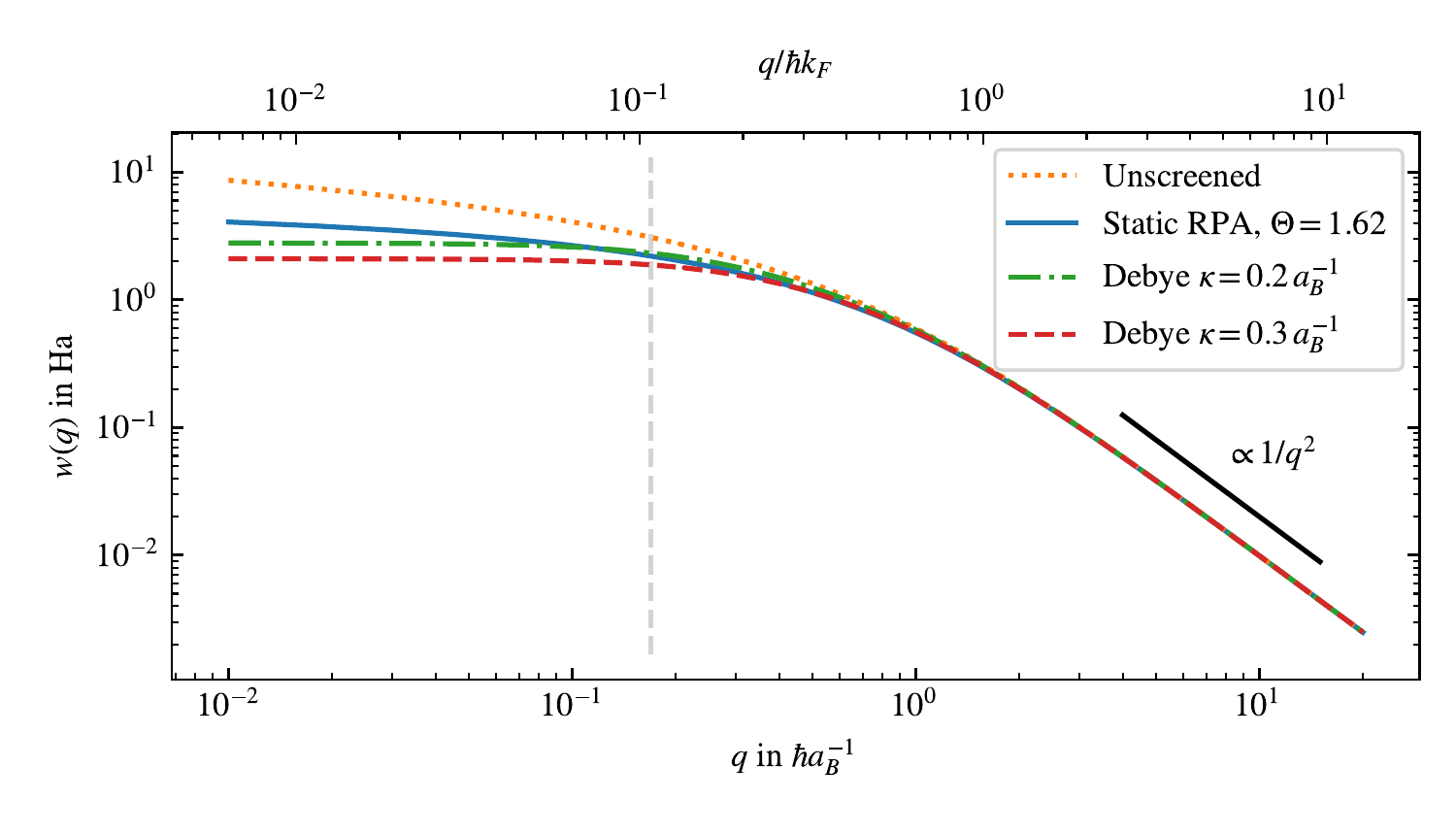}
\caption{
Fourier transform of the pair potential, Eq.~\eqref{eq:pair-potential-fourier}. Also shown is the static limit of the RPA-screened potential $w(q,\kappa=0)/\epsilon^\text{RPA}(q,\omega=0)$, where $\epsilon^\text{RPA}$ is computed using an equilibrium distribution at $r_s=0.5,\ \Theta=1.62$. The Debye-screened potential $w(q,\kappa=0.2\,\hbar a_B^{-1})$ used in this work agrees quite well with the latter. The grey dashed line depicts the $k$-point spacing used.}
\label{Fig:PairPotentialPlot}
\end{figure}

Assuming a statically screened Yukawa-type interaction with inverse screening length $\kappa$, we compute the interaction matrix elements with the wavefunctions (\ref{eq:1p-states}),
\begin{align}
   \left\langle\textbf{k}-\mathbf{q},\alpha;\textbf{p}+\mathbf{q},\beta\left| e^2\frac{e^{-\kappa r}}{r} \right|\textbf{k}\alpha;\textbf{p}\beta\right\rangle=   w\left(q\right)=-e^{2}\exp\left[(q^{2}+\kappa^{2})a^{2}\right]\text{Ei}\left(-(q^{2}+\kappa^{2})a^{2}\right),
\label{eq:pair-potential-fourier}
\end{align}
depicted in Fig.~\ref{Fig:PairPotentialPlot},
where we already chose the states in such a way that they are compatible with Eq.~\eqref{eq:interactForm}. Here, $\text{Ei}$ is the exponential integral function.
It should be noted that the limit $a\to 0$ does not exist, as then the pair potential diverges for all $q$. A finite column width is thus not only physically realistic, but also a mathematical necessity.
In our calculations we chose $a=1\,a_B=2\,r_s\,a_B$ and $\kappa^{-1}=5\,a_B=10\,r_s\,a_B$. The confinement length corresponds to $\hbar\omega=2\,\text{Ha}\approx 54.4\,\text{eV}$. The assumption of the oscillator ground state restricts temperatures to well below 2 Ha.
One physical realization of such a confinement would be a magnetic field along the axis. The above conditions would be satisfied for $B\approx 5.44\cdot 10^5\, T$. These are parameters observed in the atmosphere of neutron stars but also not far from the B-field generated that is expected to be produced in magnetized target fusion experiments at Sandia National Laboratory [\onlinecite{maglif_prl_14}].

Let us briefly comment on the choice of the screening parameter which is known to have a strong influence on the relaxation dynamics in the second order Born approximation. 
The chosen value of $\kappa$ is based on the long wavelength limit of the static polarization function in the random phase approximation (RPA), e.g.~[\onlinecite{bonitz_qkt}] which is illustrated in Fig.~\ref{Fig:PairPotentialPlot}. The underlying distribution function for the screening is the equilibrium distribution. A more accurate description of the screening can be achieved by using time-dependent nonequilibrium distribution functions, which on the other hand breaks self-consistency and energy conservation.
The dependence of the matrix elements of the Coulomb potential $w(q)$ on $\kappa$ is depicted in Fig.~\ref{Fig:PairPotentialPlot}. Preliminary results using the GW approximation where the screening parameter is established selfconsistently, confirm the choice made above.

\subsection{Stopping model}
In the following we will investigate the energy exchange between two plasma components with the momentum distribution functions $f_a(\textbf{p},t)= \pm i\hbar G^<_{\mathbf{p}\alpha}(t)$. At the initial moment, $t=t_0$, one component (the ``target'') is prepared in a thermal equilibrium state,  which is initialized self-consistently on the Hartree--Fock level
\begin{align} 
       f_a(\textbf{p},t_0)=
       \left[\exp\left(\frac{h^{\text{HF}}_{\mathbf{p}\alpha}(t_{0})-\mu^\alpha}{k_{B}T^{\alpha}}\right)+1\right]^{-1}\,,
\end{align}
where  $h^{\text{HF}}_{\mathbf{p}\alpha}(t_{0})$ is the the Hartree--Fock Hamiltonian,  Eq.~\eqref{eq:HFHamiltonianMom}, and $\mu^\alpha$ and $T^{\alpha}$ denote the chemical potential and temperature of species ``a''. The chemical potential is adjusted such that it yields the desired particle density.
The second component (the ``beam'') is given by a significantly narrower Gaussian distribution that is displaced with respect to the origin by a momentum $\textbf{p}_0$:
\begin{align}
    f^{\text{beam}}_a(\textbf{p},t_0)=
    A\exp\left[-\frac{1}{2}\left(\frac{\textbf{p}-\textbf{p}_{0}}{\sigma}\right)^{2}\right],
\end{align}
where $A$ is the amplitude (normalization constant) and $\sigma$ is the width of the beam, which is linked to the beam temperature. Starting from this initial nonequilibrium configuration of the plasma we will investigate the relaxation dynamics towards equilibrium.\\

In the following we consider two cases:
\begin{enumerate}
    \item Beam and target are the same particle species, cf. Sec.~\ref{results:SameType}. 
    \item Stopping of a beam of positive ions by an electron plasma, cf. Sec.~\ref{ss:i-in-e}.
\end{enumerate}
We will investigate the relaxation dynamics for different values of the central beam momentum, $p_0$, and mass ratio $M$. For the present test simulations we will restrict ourselves to small mass ratios, $M=1\dots 10$ and treat both species fully quantum mechanically.

\section{Numerical results}\label{s:results}
\subsection{Simulation parameters and convergence tests}\label{ss:tests}

\begin{figure}[h] 
\centering
\includegraphics[width=0.85\linewidth]{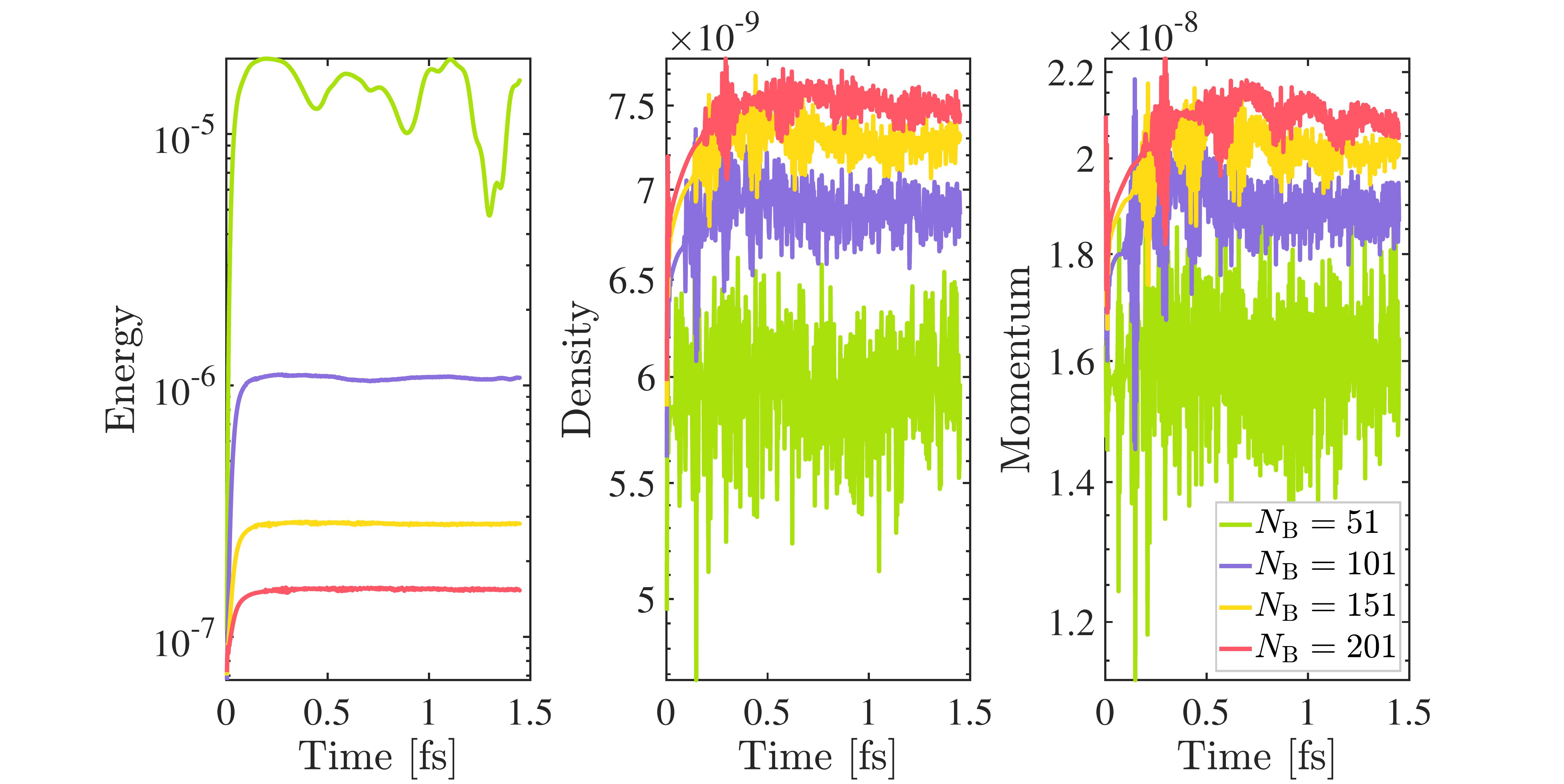}
\caption{Convergence tests: time evolution of the relative error of conserved quantities upon variation of the basis size $N_x$. The cut-off momentum  $p_\text{co}=\pm 17\,a_B^{-1}$ is kept constant, the momenta on the grid are chosen equidistantly. Based on this analysis, in all subsequent calculations we use $N_x=201.$ Two-component simulations with $N_x=201$ and $30,000$ time steps using an RK4 stepper, required $14$\,GB of RAM and took about $20$ hours on a computer node running 2x Intel Xeon Gold 6130 (Skylake), 32 cores (2.1GHz). From this, the computational demand can easily extrapolated using Tab. \ref{tab:SOAScalings}.}
\label{fig:ConvergenceTest}
\end{figure}
We use conserved quantities to benchmark the accuracy of the simulations and test the convergence with respect to the different simulation parameters such as the number of grid points $N_x$ and maximum (cut-off) momentum $p_{co}$. In a system without explicit time-dependence or external fields the total energy, the particle number density, and the total momentum are conserved. Since the equations must be solved on a finite $\mathbf{k}$-point grid, we have three convergence parameters: the cutoff wave number $p_\text{co}$ and the number of $\mathbf{k}$ points $N_x$ (we use an equidistant grid) on the range $[-p_\text{co},p_\text{co}],$. The third convergence parameter is the time step length.

In Fig. \ref{fig:ConvergenceTest} the convergence with respect to the $\mathbf{k}$-point number (basis size) is demonstrated. We use the nonequilibrium setup from Sec. \ref{results:SameType} for an ion relaxation with $m_i=5\,m_e,$ also seen in Fig. \ref{fig:dist_ionion_momentum}. The G1--G2 equations were solved with a standard fourth-order Runge-Kutta scheme with a time step of $dt=0.002\,a.u.\approx0.048\,\text{as}$, which provided converged results. The largest basis, consisting of $N_x=201$ \textbf{k}-points, fulfills the conservation laws very well: after $30,000$ time steps, the accumulated relative errors in total energy do not exceed $0.00002\%$. The errors for the density and momentum are even smaller. These parameters are, therefore, used for the calculations in the following subsections. 

The SOA equations of motion contain a momentum integration only in the collision integral defining the 1-particle dynamics. The complexity of this calculation is the same as that of the $\mathcal{G}$ propagation, where all momentum components are propagated essentially independently. Techniques that are commonly used in conventional GKBA- or two-time calculations, such as the FFT, are not necessary here and do not accelerate the calculation. Note that the present G1--G2 simulations are severely affected by \textit{aliasing}, which have been known to appear in undamped GKBA calculations. Let us briefly discuss this problem. The commutator term of the $\mathcal{G}$ equation from Eq. \eqref{eq:2pCommutatorMomRep}, which shall be repeated here,
\begin{align}
    \mathcal{G}^{\alpha\beta}_{\mathbf{kpq}}(t)\left(h^\text{HF}_{\mathbf{k}-\mathbf{q},\alpha}(t)+h^\text{HF}_{\mathbf{p}+\mathbf{q},\beta}(t)-h^\text{HF}_{\mathbf{k},\alpha}(t)-h^\text{HF}_{\mathbf{p},\beta}(t)\right)=\vcentcolon \mathcal{G}^{\alpha\beta}_{\mathbf{kpq}}\Delta h^{\text{HF},(2)}_{\mathbf{kpq},\alpha\beta}\,,
\end{align}
induces contributions to $\mathcal{G}(t)$ of the form $\sim \exp\left(\frac{1}{\mathrm{i}\hbar}\Delta h^{\text{HF},(2)}_{\mathbf{kpq},\alpha\beta}(t-t_0)\right)$. These phase factors are dependent on $\mathbf{k},\mathbf{p}$ and $\mathbf{q}.$ Hence in the collision integral, after some simulation time, a rapidly oscillating integrand appears. If the resolution of momentum space is not sufficient to resolve these oscillations correctly, the discretized integral becomes erroneous, which is a form of aliasing. We will discuss the practical occurrence and a possible solution to it in Sec.~\ref{sssec:LHF}.

\subsection{Simulation results}
In this section relaxation results for different beam and target configurations are presented. The dynamics are analyzed by investigating the time dependence of the distribution functions, their time derivatives and the mean kinetic energies per particle. 

We consider quasi-1D plasmas with a moderate Coulomb coupling,  $r_s=0.5$. A strong transverse confinenemt is imposed corresponding to an effective wire radius of $1\,a_B$, cf. Fig.~\ref{Fig:CylinderModel}, that corresponds to a 3D density of $2.15\times 10^{24}\,cm^{-3}.$ This, together with temperatures between $157,800\,K$ and $630,000\,K$, lies well within the Warm Dense Matter range, in particular these parameters are expected to be on the ICF capsule implosion path [\onlinecite{hu_ICF}] and are also achievable in magnetic target fusion devices, e.g. Ref.~\onlinecite{maglif_prl_14}. The plasma period, which is also the time scale on which correlations evolve, cf. Refs.~[\onlinecite{bonitz-etal.96pla, bonitz96pla}], is given by $\omega_{pl}=82.7\,\text{fs}^{-1}$. In our calculations we therefore focus on such ultrashort time scales.

\subsubsection{Beam thermalization in a one-component plasma}\label{results:SameType}
\begin{figure}[ht] 
\centering
\includegraphics[width=0.7\linewidth]{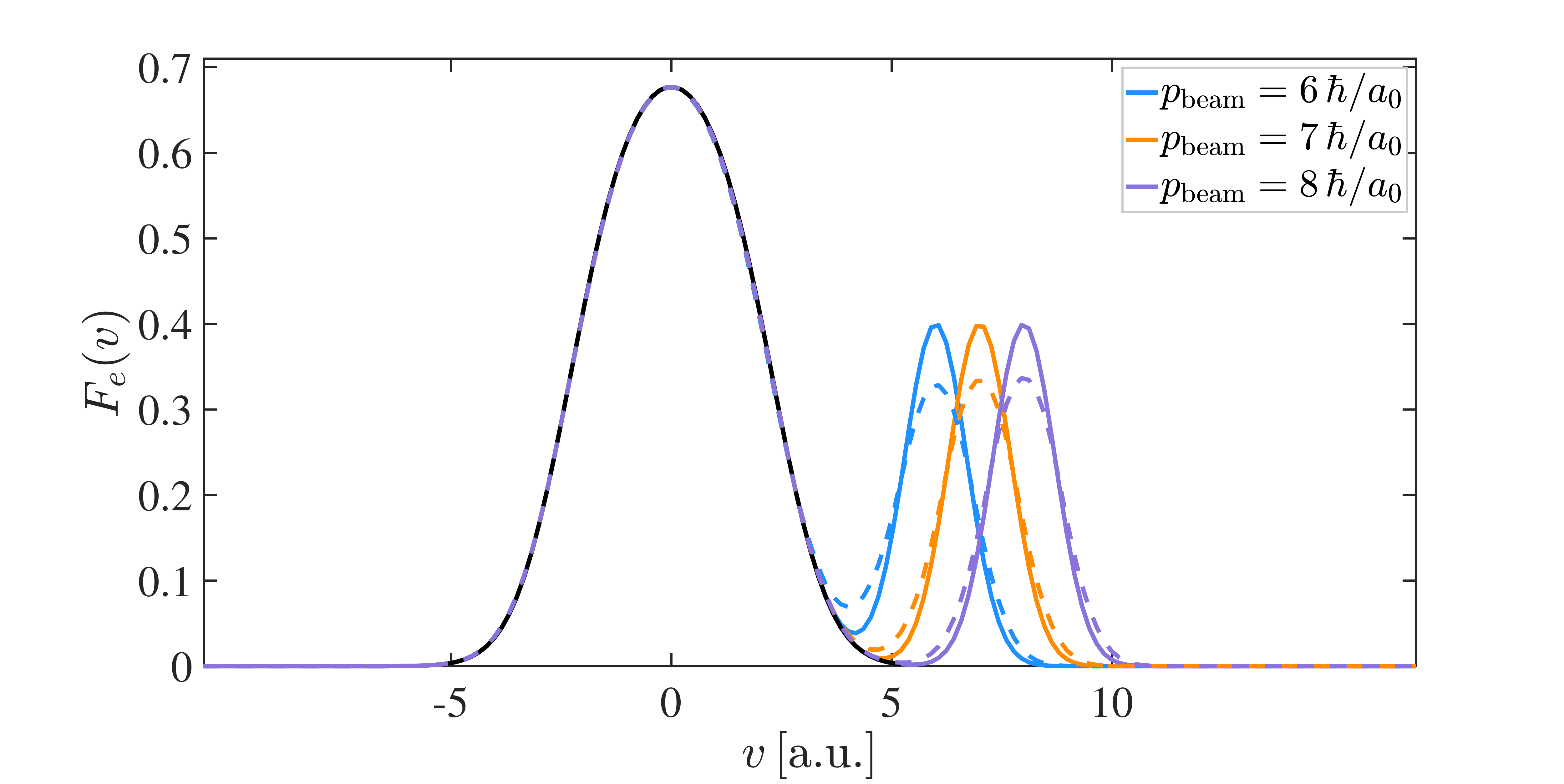}
\caption{
Time evolution of the velocity distribution function of a pure electron plasma. The ``target'' electrons are characterized by $r_s=0.5$ and $\Theta=1.62$ and scatter with and electron beam of a fixed shape but different mean momenta $p_{\rm beam}$ indicated in the figure. Solid lines:  initial time, dashed lines: distributions at $t=1.5\,\text{fs}.$ Little equilibration is visible, especially for higher beam velocities. 
}
\label{fig:velocity_electronbulk_electronbeam}
\end{figure}

All figures in this subsection present simulation data where target and projectile particles are of the same type. In Fig. \ref{fig:velocity_electronbulk_electronbeam} the relaxation process of electron projectiles with electrons of a target plasma is shown for different projectile velocities. The plasma parameters of the target were chosen to be $r_s=0.5$ and $\Theta=1.62$. The Gaussian distributions representing the projectiles have the amplitude $0.4$ (maximum occupation number at center momentum) and variance $\sigma^2=0.5\,\text{Ha},$ corresponding to an effective beam temperature of $157,900\,\text{K}$ (for $m=m_e$). The ratio between beam and target densities is $n_{beam}/n_{plasma}\approx 0.23,$ which implies that these calculations are well beyond linear response. 
Figure~\ref{fig:velocity_electronbulk_electronbeam} shows that only little relaxation has occurred over the simulation duration of $\sim 1.5\,\text{fs}$. We also observe that the relaxation speed is decreased when the projectile velocity increases. An explanation for this trend is given in Sec.~\ref{s:discussion}.

\begin{figure} 
\centering
\includegraphics[width=0.7\linewidth]{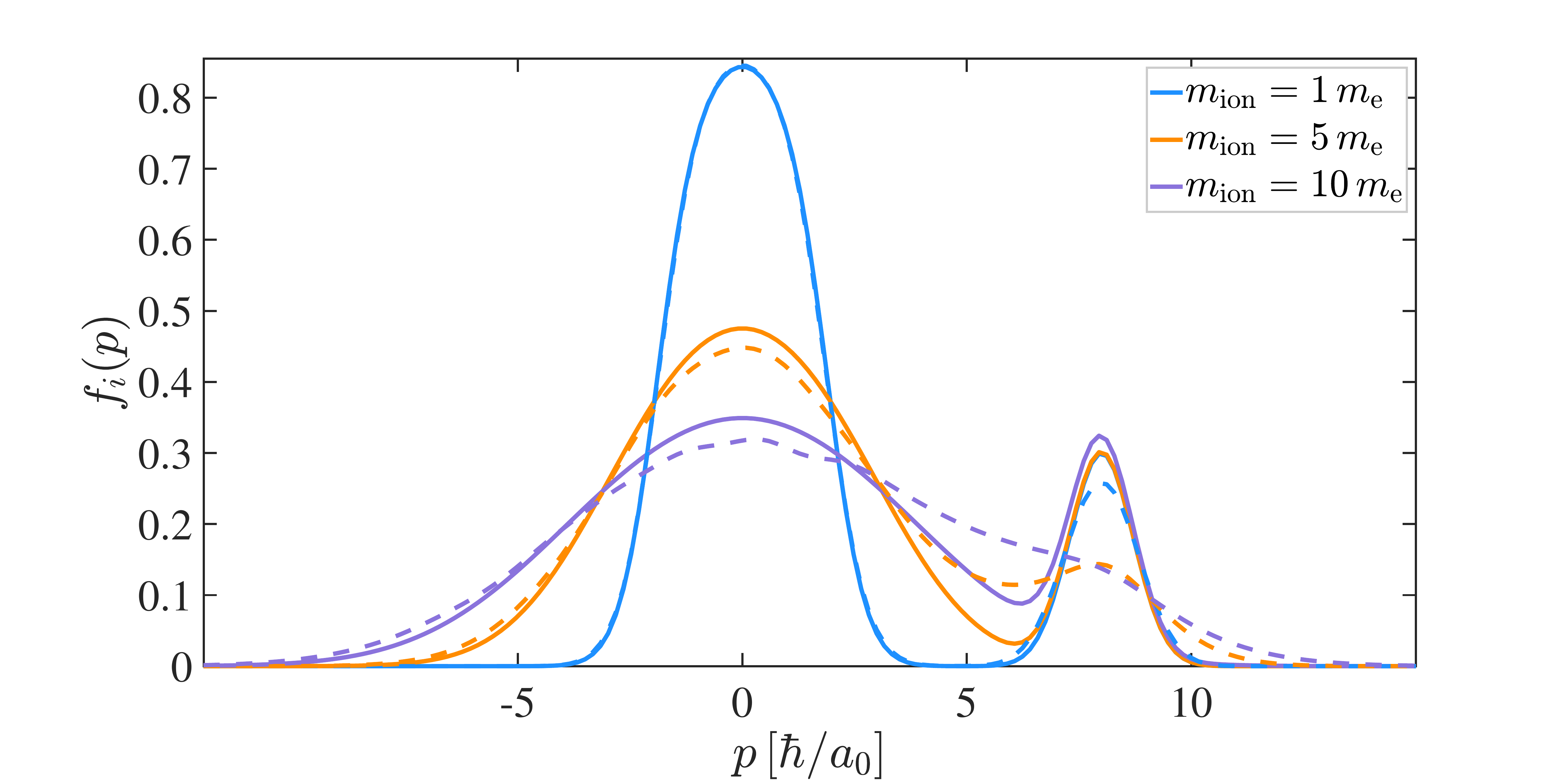}
\caption{
Time evolution of the distribution function in a one-component plasma with a ``bump on tail'' distribution. Initial parameters: $r_s=0.5$ and $T=157,900\,\text{K}$. The shape of the initial beam distribution is fixed. Different colors correspond to particles of different mass. Significantly more equilibration is visible the higher the masses of the particles are.}
\label{fig:dist_ionion_momentum}
\end{figure}

In the next series of simulations, shown in Fig. \ref{fig:dist_ionion_momentum}, the projectile momenta were fixed. Instead, the masses of all particles were varied at $p_B=8\hbar/a_0$. Here, the target temperature is fixed at $315,800\,\text{K}$, for all calculations shown. The fixed width of the Gaussian, $\sim \exp\left(-\left[\mathbf{p}-\mathbf{p}_B\right]^2/2m_i k_BT_B\right),$ where $\sigma^2=k_B T_B m_i /m_e$ is chosen to be constant, which corresponds to temperatures of $157,900\,\text{K}$ (for $m_i=1m_e$), $31,600\,\text{K}$ ($5m_e$) and $15,800\,\text{K}$ ($10m_e$), respectively. The computations shown in Fig. \ref{fig:dist_ionion_momentum} demonstrate that equilibration is faster when the particle mass increases.

In the following we provide an analytical explanation for this observation. Analyzing the results we conclude that only particles of identical or near-identical velocity effectively scatter and exhibit a significant exchange of momentum and energy. This is, of course, a specific of the quasi-1D geometry. Mathematically, this can be understood from the memory form of the collision integral in second Born approximation. In $d$ dimensions, we have [\onlinecite{bonitz_qkt}],
\begin{align}
    I_{\mathbf{p}\alpha}(t)&=\frac{2}{\hbar}\sum\limits_{\beta} \int_{0}^{t-t_0}\text{d}\tau\int\frac{\text{d}\mathbf{p}_2}{(2\pi\hbar)^d}\int\frac{\text{d}\mathbf{q}}{(2\pi\hbar)^d}w_{\mathbf{q}}\left[w_\mathbf{q}\pm \delta_{\alpha\beta}w_{\mathbf{p}-\mathbf{p}_2-\mathbf{q}}\right]\cos\left[\frac{h^{\text{HF}}_{\mathbf{p},\alpha}+h^{\text{HF}}_{\mathbf{p}_2,\beta}-h^{\text{HF}}_{\mathbf{p}+\mathbf{q},\alpha}-h^{\text{HF}}_{\mathbf{p}_2-\mathbf{q},\beta}}{\hbar}\tau\right]\nonumber\\
    &\times \left[G^<_{\mathbf{p}+\mathbf{q},\alpha}(t-\tau)\,G^<_{\mathbf{p}_2-\mathbf{q},\beta}(t-\tau)\,G^>_{\mathbf{p},\alpha}(t-\tau)\,G^>_{\mathbf{p}_2,\beta}(t-\tau)-G^>_{\mathbf{p}+\mathbf{q},\alpha}(t-\tau)\,G^>_{\mathbf{p}_2-\mathbf{q},\beta}(t-\tau)\,G^<_{\mathbf{p},\alpha}(t-\tau)\,G^<_{\mathbf{p}_2,\beta}(t-\tau)\right]\,.
\end{align}
The exchange energy in these calculations is negligible,  compared to the kinetic energies, so we approximate  $h^\text{HF}_{\mathbf{p},\alpha} \approx p^2/2m_\alpha$. For an analysis of the dominant contributions to the collision integral, we consider the Markov limit, $G^\gtrless_{\mathbf{p},\alpha}(t-\tau)\approx G^\gtrless_{\mathbf{p},\alpha}(t)$, together with the limit $t_0\to -\infty$, (weakening of initial correlations [\onlinecite{bonitz_qkt}]), which yields
\begin{align}
    I_{\mathbf{p}\alpha}(t)&=\frac{2}{\hbar}\sum\limits_{\beta} 
    \int\frac{\text{d}\mathbf{p}_2}{(2\pi\hbar)^d}\int\frac{\text{d}\mathbf{q}}{(2\pi\hbar)^d}w_{\mathbf{q}}\left[w_\mathbf{q}\pm \delta_{\alpha\beta}w_{\mathbf{p}-\mathbf{p}_2-\mathbf{q}}\right]\delta\left[\mathbf{q}\cdot\left(\mathbf{v}_2-\mathbf{v}\right)+\frac{m_\alpha+m_\beta}{m_\alpha m_\beta}q^2\right]\nonumber\\
    &\times \left[G^<_{\mathbf{p}+\mathbf{q},\alpha}(t)\,G^<_{\mathbf{p}_2-\mathbf{q},\beta}(t)\,G^>_{\mathbf{p},\alpha}(t)\,G^>_{\mathbf{p}_2,\beta}(t)-G^>_{\mathbf{p}+\mathbf{q},\alpha}(t)\,G^>_{\mathbf{p}_2-\mathbf{q},\beta}(t)\,G^<_{\mathbf{p},\alpha}(t)\,G^<_{\mathbf{p}_2,\beta}(t)\right]\,,\label{eq:MarkovCollInt}
\end{align}
where the velocity $\mathbf{v}=\mathbf{p}/m_\alpha$ has been introduced. Since $w$ 
has its dominant contributions at $q\lesssim 0.5 a_B$, cf. Fig.~\ref{Fig:PairPotentialPlot},
it is reasonable to consider the small-$\mathbf{q}$ limit, where the argument of the $\delta$-function becomes $\mathbf{q}\cdot(\mathbf{v}_2-\mathbf{v}).$ Now there are two cases, where the argument is $0$, and momentum between particles is exchanged efficiently during a collision: first, if the transferred momentum $\mathbf{q}$ is perpendicular to the velocity difference $\mathbf{v}_2-\mathbf{v},$ and, second, if the velocities are equal, $\mathbf{v}_2=\mathbf{v}.$ The first case can be excluded in a quasi-1D geometry. The primary condition for collisions is thus a resonance -- the equality of the velocities of the two scattering partners. Thus, in the Markov limit only ``on-shell'' scattering contributions are relevant for the relaxation. Due to the strongly reduced phase space this is possible only if target particles with a velocity close to the beam velocity exist.

In contrast, in the full non-Markovian case, such as in the G1--G2 calculations, where the difference $t-t_0$ is finite, the $\delta$-function is significantly broadened to a degree that ``off-shell''-scattering processes are relevant where the kinetic energy is not strictly conserved. Nevertheless, the dominant scattering contributions still originate from particles with resonant velocities which underlines the relevance of the above analytical analysis.\\

\begin{figure} 
\centering
\includegraphics[width=0.7\linewidth]{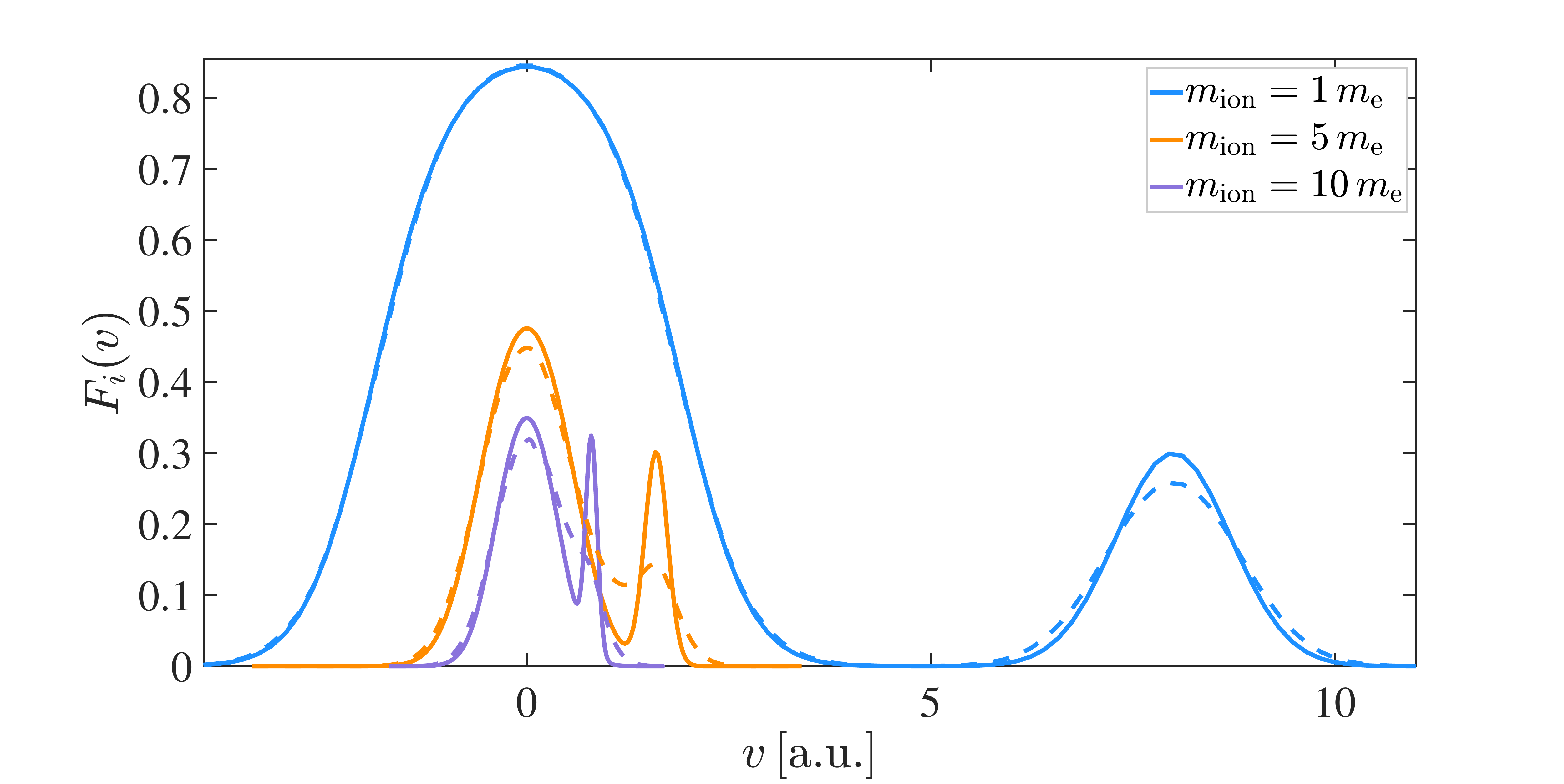}
\caption{
Same data as in Fig. \ref{fig:dist_ionion_momentum}, but on a velocity scale, which is the relevant scale in the explanation of the findings, cf. Sec.~\ref{results:SameType}, in particular Eq.~\eqref{eq:MarkovCollInt} and the discussion thereafter. Note that the density of states in $v$-representation depends on the mass of the particle (linear in mass $m_{e,i}$), which has to be taken into account when trying to compute observables from this representation.
}
\label{fig:dist_ionion_velocity}
\end{figure}

Figure \ref{fig:dist_ionion_velocity} illustrates how with larger particle masses and similar momentum distribution their velocity scale shrinks and therefore collisions become more likely according to the analysis above. As a consequence, the thermalization proceeds much faster, for $M=10$, compared to $M=5$ and $M=1$. The analysis above also applies to two-component systems, where the connection between $v_{e/i}$ and $p_{e/i}$, given by a rescaling by mass $m_{e/i}$, is not identical for the two species. Such systems are investigated in Sec.~\ref{ss:i-in-e}, where beam and target velocities are chosen in such a way that they overlap.

\clearpage

\begin{figure}[b] 
\centering
\includegraphics[width=0.7\linewidth]{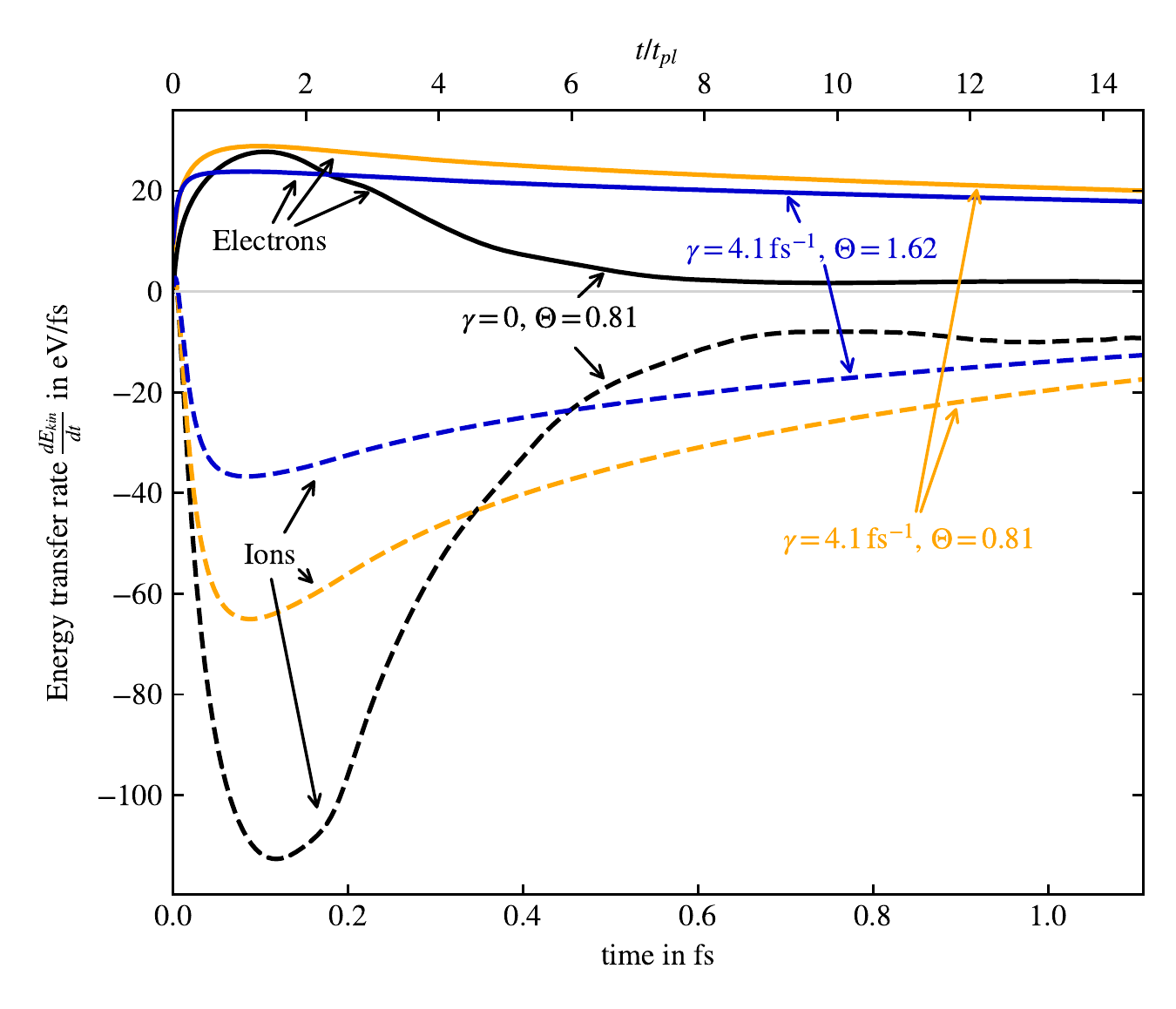}
\caption{
Kinetic energy change of electrons (full lines) and ions (dashed lines) as a function of time for the simulations of Fig. \ref{fig:vdist_etarget_iprojectile}. Black: HF-GKBA (no damping), blue (yellow): LHF-GKBA for $\Theta=0.81 $  $(\Theta=1.62)$. 
}
\label{fig:tdstopping}
\end{figure}

\subsubsection{Ion stopping in an electron plasma. Aliasing effects}\label{ss:i-in-e}
We now consider stopping of an ion beam by an initially thermal electron plasma.
In accordance with Sec.~\ref{results:SameType}, we focus on situations where the velocity distributions of different particle species overlap. The results are presented in Figs.~\ref{fig:tdstopping} and \ref{fig:vdist_etarget_iprojectile}.

Consider first the black line in Fig.~\ref{fig:vdist_etarget_iprojectile} that shows the time-dependent distribution function (left column) and its time-derivative (right column) for a G1--G2 simulation based on the HF-GKBA of an electron target given by the parameters $r_s=0.5,\ \Theta=0.81$. The electrons are impacted by an ion beam with mass $M=5$ and a slightly lower density as compared to the previous section. While the ion distribution broadens in time, due to the scattering with the electrons, the latter display a non-trivial dynamic: apparently, a distribution with two maxima emerges and becomes stationary, where the second maximum is aligned with the velocity of the ions. However, a closer analysis of the derivative $dF_e/dt$ reveals that this is a numerical artifact. The time derivative assumes large values that oscillate both with respect to momentum and time for which there is no physical reason. Due to these oscillations the thermalization of the electron distribution function is artificially reduced. This can further be observed in Fig.~\ref{fig:tdstopping}, where in the black curves, the change of the kinetic energy per particle is shown, which after an initial peak quickly decays, indicating stationary behaviour.

To understand the origin of this unexpected behavior we varied the time step and the number of k-points and observed that this behavior does not change significantly: merely the time after which stationarity occurs is prolonged for larger numbers of k-points. The conclusion is that this behavior is a consequence of aliasing that was mentioned above which has an especially drastic effect in 1D. It will be further explained in the next section, where we also present a practical solution that can be applied within the G1--G2 scheme.

\subsubsection{Reduction of aliasing by means of the Lorentzian HF-GKBA}\label{sssec:LHF}

Aliasing is a type of error that has its origin in the discretization of originally continuous data with dense oscillations. In our case, the momentum space is discretized, and $\mathcal{G}$ contains oscillatory contributions of the type $\sim \exp\left(\frac{1}{\mathrm{i}\hbar}\Delta h^{\text{HF},(2)}_{\mathbf{kpq},\alpha\beta}(t-t_0)\right)$. These become more and more dense, as $t-t_0$ grows, which immediately affects the one-particle  collision integral that is a trace over $w_q\mathcal{G}^{\alpha\beta}_{\mathbf{kpq}}(t)$. The practical realization in a simulation is always based on a discretization. If the oscillation density is on the scale of the discretization spacing and above, the integrand is not resolved sufficiently well, and the approximate integration on the grid becomes erroneous. Since a simple integral can be considered the $0$-component of the Fourier transform, this effect can be seen as a form of aliasing well-known from the spectral analysis of discretized signals. In particular, in accordance with Nyquist's theorem, aliasing occurs earlier, the coarser the momentum grid is. A discussion of aliasing in 1D, 2D and 3D simulations of uniform systems using standard GKBA and the G1--G2 scheme is given in Ref.~[\onlinecite{makait_msc_2022}].

In the G1--G2 framework the aliasing problem can be reduced by including correlations in the single-particle propagation  approximately, by using the LHF-GKBA, cf. Eq.~\eqref{eq:hhf-damped}, which yields a change in the two-particle commutator,
\begin{align}
    \left[ h^{\text{HF},(2)}(t),\mathcal{G}(t) \right]_{\mathbf{kpq}}^{\alpha\beta}=\mathcal{G}^{\alpha\beta}_{\mathbf{kpq}}\left(h^\text{HF}_{\mathbf{k}-\mathbf{q},\alpha}(t)+h^\text{HF}_{\mathbf{p}+\mathbf{q},\beta}(t)-h^\text{HF}_{\mathbf{k},\alpha}(t)-h^\text{HF}_{\mathbf{p},\beta}(t)-4\mathrm{i}\hbar\gamma \right)\,.
\end{align}
Here, $\gamma$ is a real parameter, which gradually damps out contributions from the past. With this, the dense oscillations mentioned above now assume the form $\sim \exp\left(\frac{1}{\mathrm{i}\hbar}\Delta h^{\text{HF},(2)}_{\mathbf{kpq},\alpha\beta}(t-t_0)-4\gamma(t-t_0)\right)$. The formerly critical regions of large $t-t_0$ are now damped. With sufficiently large $\gamma,$ aliasing can be reduced to unnoticeable levels.

However, since a finite $\gamma$ corresponds to simplified correlation dynamics, it must not be chosen too large. In particular, if the time scale of damping is shorter than that of correlations (correlation time [\onlinecite{bonitz-etal.96pla}]), i.e. $\gamma\gg \omega_{pl},$ the total energy will not be conserved, as relevant contributions might be damped out too fast. In high-density 3D systems a second-order limit has been derived, cf. Ref.~[\onlinecite{Bonitz1999}], given by
\begin{align}
    \gamma_{3D}=\frac{1}{3^{2/3}\pi^{5/6}}\sqrt{\frac{me^2}{\hbar^2 \epsilon_0}}\,n^{-1/6}_{3D}\omega_{pl}\,.
\end{align}

Here, $n_{3D}$ is the 3D density, which we can estimate, for our 1D system, by $n_{3D}=n_{1D}/\pi a^2$, where $a$ is the wire radius. At $r_s=0.5$, this yields $\omega_{pl}=82.7\,\text{fs}^{-1}$ and $\gamma_{3D}=65.7\,\text{fs}^{-1}.$ Since this value is derived from qualitative arguments only 
we choose a significantly smaller value for tests and set  $\gamma=4.1\,\text{fs}^{-1}.$ Our calculations show that the relative total energy conservation violation is below $10^{-4}$. This means, even though the LHF model is very rough, the present choice of $\gamma,$ does not overestimate the quasiparticle damping.

\begin{figure}[h] 
\centering
\includegraphics[width=0.9\linewidth]{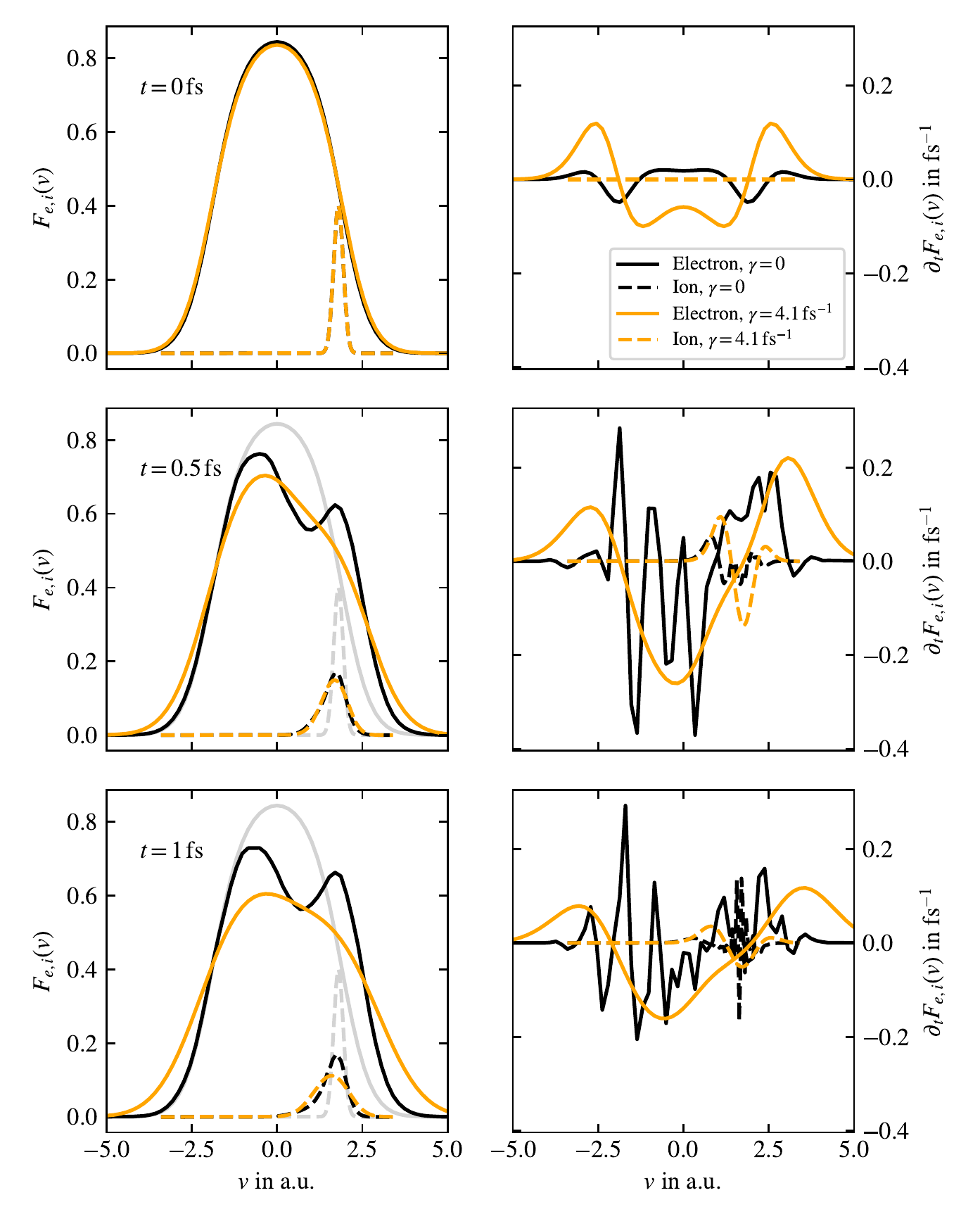}
\caption{Time evolution (from top to bottom) of the velocity distributions (left column) and their time derivatives (right column) of initially thermal electrons ($r_s=0.5,\,\Theta=0.81$) subject to an ion beam with $m_i=5\,m_e$ with central beam momentum of $9\hbar/a_B$.
Time $t=0$ corresponds to the end of the adiabatic switch on of the pair interaction and the point of time when the ions are added into the system. 
We compare G1--G2 simulations with Hartree-Fock propagators (HF-GKBA, $\gamma=0$) and with exponentially damped propagators (LHF-GKBA, $\gamma=4.1/$fs.), cf. Eq.~\eqref{eq:hhf-damped}, respectively. The grey lines on the left represent the distributions at $t=0$.
}
\label{fig:vdist_etarget_iprojectile}
\end{figure}

We now repeat the simulations of Fig.~\ref{fig:vdist_etarget_iprojectile} with the finite $\gamma$ and observe dramatic changes.
The orange curves in Fig.~\ref{fig:vdist_etarget_iprojectile} depict the corresponding results for the time-dependent distribution function and its time-derivative using the LHF: In contrast to the undamped HF-GKBA, the time-derivative does not contain oscillations. Instead, the ion distribution rises more `on the left' of its initial peak than `on the right', indicating that they continue being stopped by the electronic target. The electronic derivative shows the complementary dynamics, i.e. electron acceleration in positive direction. Fig.~\ref{fig:tdstopping} confirms (see the orange curves) that the energy exchange between ions and electrons no extends to much longer times than in case of undamped propagators. 

Finally, we consider another case, where the electron temperature $\Theta$ is increased by a factor $2$. This case is included in Fig.~\ref{fig:tdstopping} as well, cf. the blue curves. Here, the energy exchange is reduced in comparison to the colder system, which is due to the less steep slope of the target distribution, which reduces the drift of the projectile distribution and therefore the net energy exchange. At the same time the dynamics of the distribution functions is only weakly altered, but differences are visible in the time derivatives.

\subsubsection{Ion stopping in an electron plasma. Influence of the mass ratio}\label{ss:i-in-e-2}
\begin{figure}
    \centering
    \includegraphics[width=0.9\textwidth]{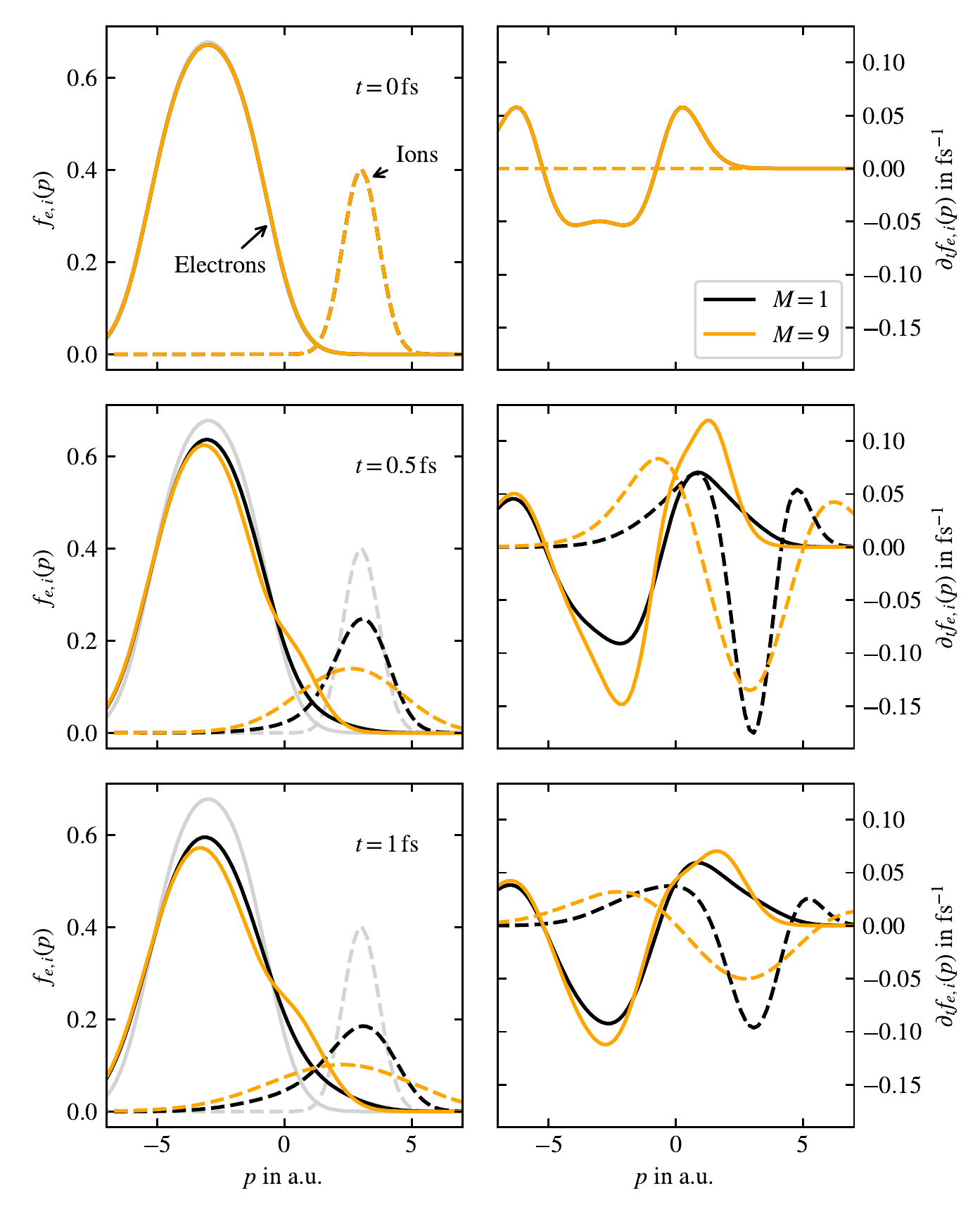}
    \caption{Equilibration of symmetrically displaced distributions for two different ion masses $M$. The target electron distribution is given by $r_s=0.5$, $\Theta=1.62$, but centered around $p=-3\,\hbar a_B^{-1}.$ The grey lines on the left represent the $t=0$ distributions.}
    \label{fig:ei_Masseffect}
\end{figure}

In our final series of calculations, we study the influence of the mass ratio $M$. In particular, we simulate a two-component system where the two components are displaced in $\mathbf{p}$ space in opposite directions, to $\pm\mathbf{p}.$ This situation can be achieved in laboratory by quickly accelerating carriers of opposite charges distributed around $\mathbf{p}=0$ (electrons and holes, or electrons and single-charge cations) in an external electric field. In accordance with the previous section we use the LHF-GKBA with $\gamma=4.1\,\text{fs}^{-1}$ to reduce the aliasing errors. Fig.~\ref{fig:ei_Masseffect} shows the relaxation for $M=1$ and $M=9.$ The electron part of the plasma is characterized by $r_s=0.5$ and $\Theta=1.62$ centered around $\mathbf{p}=-3\,\hbar a_B^{-1}$, whereas the ion part is given by the Gaussian used in the prior sections, just centered around $\mathbf{p}=3\,\hbar a_B^{-1}.$ 

It is evident that the equilibration happens faster if the ions are heavier. This is partially due to much more effective ion-ion collisions, since identical initial momentum distributions but higher mass implies a smaller velocity range. This in turn leads to more efficient collisions. 

The initial broadening due to ion-ion collisions also changes the interaction efficiency between ions and electrons. In velocity space, the distribution function of the heavier ions is nearer to $0$ than that of the lightweight ions, giving a stronger overlap with the electronic counterpart. This is even increased after the initial broadening, which is more effective the heavier the ion is.

\clearpage

\section{Conclusions and outlook}\label{s:discussion}

In summary, we have studied the the performance of the G1--G2 scheme in a quasi-one-dimensional two-component stopping setup. It was found to be very stable and efficient over long simulation times. Equilibration between the two particle species could only be observed if the velocities of the colliding particles is comparable. This rather strict condition could be derived analytically from the conservation of single-particle energies in the Markov limit and is a result of the strongly reduced quasi-1D phase space. Furthermore, it turned out that G1--G2 calculations are not devoid of aliasing, which has an especially strong effect in one dimension. One way to combat aliasing is a correlated GKBA such as the LHF-GKBA which can be introduced into the G1--G2 scheme in a straightforward manner. Choosing a small damping, $\gamma \ll 1/\omega_{pl}$, provides a satisfactory solution that does not violate conservation laws.\\

While our method so far has only been used with test parameters, in particular small ion masses, $m_i \le 10 m_e$, it can be straightforwardly extended to realistic physical systems such as quantum plasmas in strong magnetic fields with real ions. The latter can also be treated classically, as explained in Ref.~\onlinecite{rinton-book} which will allow one to further simplifies the simulations. 
Moreover, future investigations will involve more sophisticated selfenergy approximations, such as $GW$ and DSL which do not pose a significant additional challenge within the G1-G2 scheme, as shown in Ref.~\onlinecite{joost_prb_22}. These approximations have the capability to capture important physical effects such as beam-plasma instabilities, and the energy transfer might be enhanced due to the appearance of acoustic plasmons. We expect that systematic parameter scans will be possible with our scheme and will allow the computation of the stopping power as a function of the beam velocity $v$ for various plasma parameters $r_s,\,\Theta$. These calculations are not limited to linear response: dense particle beams or nonequilibrium targets are also within reach. 
We also note that the results for the stopping power in an electron-ion plasma are closely related to the temperature relaxation which is presently of high interest in warm dense matter, e.g. Refs. \onlinecite{ohde-etal.96pp,vorberger_pre_10, daligault_pre_19}. We expect that our simulations will approach, in the long-time limit, the stage of a two-temperature quasi-equilibrium plasma and will yield improved results for the equilibration rates.

Even though present hardware restrictions limit G1-G2 simulations to a quasi-1D geometry, the results are expected to be useful for a better understanding of the energy exchange in scattering processes in confined geometries. Examples include thermalization in strongly magnetized stellar objects or ion beam plasma heating in highly compressed matter in a strong magnetic field, including direct drive ICF and magnetized target fusion.

\section*{Acknowledgements}
We acknowledge fruitful discussions with J.-P. Joost and N. Schlünzen. This work has been supported by the Deutsche Forschungsgemeinschaft via grant BO1366/16.

\end{document}